\documentclass{JHEP3}
\usepackage{amsmath}
\usepackage[numbers]{natbib}
\usepackage{graphicx}
\usepackage[bf,small]{caption2}
\setcaptionwidth{.9\textwidth}
\DeclareMathOperator{\Tr}{Tr}
\DeclareMathOperator{\tr}{tr}
\newcommand{\Ukk}{u_\Lambda}
\newcommand{\del}{\partial}

\makeatletter
\renewcommand\@ENVwarn[1]{}
\makeatother

\title{Rho meson condensation at finite isospin chemical potential
in a holographic model for QCD}

\author{Ofer Aharony{}$^1$, Kasper Peeters{}$^2$, Jacob
  Sonnenschein{}$^{3,4}$ and Marija Zamaklar{}$^5$\\
\llap{{}$^1$}Department of Particle Physics, Weizmann Institute of
Science, Rehovot 76100, Israel.\\
\llap{{}$^2$}Institute for Theoretical Physics, Utrecht University, P.O.~Box 80.195,
3508 TD Utrecht, The Netherlands.\\
\llap{{}$^3$}School of Physics and Astronomy,
The Raymond and Beverly Sackler Faculty of Exact Sciences,
Tel Aviv University,
Ramat Aviv, 69978, Israel.\\
\llap{{}$^4$}Albert Einstein Minerva Center, Weizmann Institute of Science, Rehovot 76100, Israel.\\
\llap{{}$^5$}Department of Mathematical Sciences,
Durham University,
South Road,
Durham DH1 3LE, United Kingdom.\\
~\\
\email{ofer.aharony@weizmann.ac.il}\\
\email{kasper.peeters@aei.mpg.de}\\
\email{cobi@post.tau.ac.il}\\
\email{marija.zamaklar@durham.ac.uk}}

\abstract{We analyze the effect of an isospin chemical potential $\mu_I$ in
  the Sakai-Sugimoto model, which is the string dual of a confining
  gauge theory related to large $N_c${} QCD, at temperatures below the chiral symmetry restoration
  temperature. For small chemical potentials we show that the results
  agree with expectations from the low-energy chiral Lagrangian, and
the charged pion condenses. When the chemical potential reaches a
critical value \mbox{$\mu_I = \mu_{\text{crit}} \simeq 1.7 m_{\rho}$}, the lowest
vector meson (the ``rho meson'') becomes massless, and it condenses (in addition
to the pion condensate)
for \mbox{$\mu_I > \mu_{\text{crit}}$.} This spontaneously breaks the
rotational symmetry, as well as a residual $U(1)$ flavor symmetry. We
numerically construct the resulting new ground state for $\mu_I > \mu_{\text{crit}}$.}

\preprint{\small DCPT-07/55, ITP-UU-07/49, TAUP-2865/07,
  WIS/16/07-SEPT-DPP, NI-07094}

\begin{document}
\section{Introduction and summary of results}

Unraveling the phase diagram of QCD, in particular at non-zero baryon
density, remains a major challenge (see~\cite{Schafer:2005ff} for a
recent review).  Although lattice simulations have provided us with a
great deal of information about the finite-temperature behavior of
QCD, it is technically much less clear how to reliably extract the
physics at finite baryon chemical potential, because of the notorious
sign problem. One way to get around this problem is to consider
non-zero isospin chemical potential
instead~\cite{Alford:1998sd,Son:2000xc}. Although there are no
physical systems in nature in which only a large isospin density is
achieved (because weak decays do not conserve isospin), studying such
systems may nevertheless provide us with useful knowledge about the
full phase diagram of QCD.\footnote{Neutron stars have a large isospin
  density, but in this case the baryon density is also large, in
  contrast to the idealized situation studied here, in which the
  chemical potential and charge densities are turned on only in a
  $U(1)$ subgroup of the $SU(2)$ isospin group.}

The effects of a small isospin chemical potential~$\mu_I$ can be
analyzed using the chiral Lagrangian~\cite{Son:2000xc}. For chemical
potentials below the mass~$m_\pi$ of the pion, one finds that the pion
masses (as well as the masses of other isospin triplets) 
split: one of them increasing, one decreasing and one remaining
constant. The ground state, however, remains unmodified from that of
QCD at $\mu_I=0$, until $\mu_I$ reaches the pion mass~$m_\pi$. At this
point, one of the charged pions condenses, with a magnitude that grows
with $\mu_I$.  This situation persists until~$\mu_I$ becomes of the
order of the mass of the first massive meson, where chiral
perturbation theory breaks down, and it is not clear how to analyze
what happens for larger values of $\mu_I$ (except by lattice
simulations, but these are hard to perform for large pion
condensates). One may consider toy models in which higher-derivative
interactions or interactions with massive vectors are added, such as
the Skyrme model, but it is not clear how to do this in a controllable
way.  Only at extremely large values of~$\mu_I$ does one regain
analytic control. Here, due to asymptotic freedom, perturbation theory
applies and predicts a~$\langle \bar{u}\gamma_5 d\rangle$ condensate,
much like a fermion superfluid~\cite{Son:2000by}. Given that the
quantum numbers of the condensate are the same as those of the pion
condensate for small $\mu_I$, it has been
conjectured~\cite{Son:2000by} that no phase transition occurs as
$\mu_I$ is increased from~$m_\pi$ to $\infty$.  However, this remains
a conjecture, and it has also been suggested that the rho meson may
condense when the isospin chemical potential is of the order of the rho
meson mass, spontaneously breaking the rotational
symmetry~\cite{Voskresensky:1997ub,Lenaghan:2001sd,Sannino:2002wp} 
(see~\cite{Sannino:2003fj} for a review). So far, lattice studies are
able to reproduce the results of the chiral
Lagrangian (see, for instance, \cite{Kogut:2002zg}), and studies are currently under way to
explore more of the intermediate and high isospin chemical potential regimes.
\medskip

In the present paper we analyze the behavior of QCD-like theories at
moderate values of the isospin chemical potential, using an
alternative analytic route.  Instead of analyzing QCD, we will analyze
the Sakai-Sugimoto model~\cite{Sakai:2004cn,Sakai:2005yt} (see also
the review~\cite{Peeters:2007ab})\footnote{Flavor chemical potentials
  in other holographic models, which do not exhibit dynamical chiral
  symmetry breaking, were recently studied
  in~\cite{Nakamura:2006xk}.}.
This model has a dimensionless parameter $\lambda_5/R$ (we will
explain the meaning of $\lambda_5$ and $R$ in the next section). It is
an easily-analyzed weakly-curved string theory in the limit of large
$\lambda_5/R$, while it goes over to large $N_c$ QCD in the opposite
limit; our analysis will be limited to the former limit, where the
model contains many additional degrees of freedom beyond those of
QCD. For any value of the dimensionless parameter, this model is dual
to a large $N_c$ confining $SU(N_c)$ gauge theory with $N_f$ massless
flavors, which exhibits spontaneous chiral symmetry breaking.  We will
analyze the case of an isospin chemical potential, as opposed to a
baryon chemical potential\footnote{A baryon chemical potential in this
  model was recently analyzed
  in~\cite{Bergman:2007wp,Rozali:2007rx,Yamada:2007ys,Davis:2007ka,Kim:2007zm}.},
for three reasons.  First, we will work in the large $N_c$ limit, and
in this limit the mesons (which are the relevant degrees of freedom
for the isospin chemical potential analysis) are much more similar to
the mesons of QCD than the baryons (which are relevant for the baryon
chemical potential analysis); in particular they are weakly
interacting and have finite masses. Note also that at large baryon
chemical potential the phase diagram for large $N_c$ is quite
different than for $N_c=3$~\cite{Shuster:1999tn}, but the phase
diagrams seem similar for the isospin chemical potential case. Second,
in the Sakai-Sugimoto model, mesons are much easier to control than
baryons. The light mesons correspond simply to modes of massless
fields living on the D8-branes, while baryons are complicated
solitonic objects with a size of order the string
scale~\cite{Nawa:2006gv,Hata:2007mb,Hong:2007kx,Hong:2007ay} (or maybe even
smaller), so it is not clear how to write down the full effective
action describing the baryons and their condensation. Third, as
mentioned above, the phase diagram for a finite isospin chemical
potential can potentially be compared with lattice simulations,
because the ``sign problem'' which plagues simulations with finite
baryon chemical potential is absent~\cite{Alford:1998sd,Son:2000xc}.

The Sakai-Sugimoto model describes a chiral gauge theory with an
$SU(N_f)_L\times SU(N_f)_R$ global symmetry which is spontaneously broken at
zero temperature. At finite temperature it exhibits (at zero chemical potential)
three possible phases~\cite{Aharony:2006da,Parnachev:2006dn}. In the
low-temperature phase, the model is confined and the chiral symmetry
is broken to its diagonal subgroup. In the intermediate-temperature
phase, which only appears for sufficiently large constituent quark mass
\footnote{The
  Sakai-Sugimoto model has another dimensionless parameter,
  which we will denote by $L/R$ in the next section, which
  determines the constituent quark mass (as measured, for instance,
  from the spectrum~\cite{Kruczenski:2004me} or decay
  rates~\cite{Peeters:2005fq} of high-spin mesons). This parameter has
  a large effect on the physics in the limit we work in, though it is
  expected to decouple in the QCD limit. The bare quark mass (and,
  consequently, also the pion mass) is zero
  for any value of this parameter. Some recent attempts to add
  non-zero bare quark masses to this model appear in
  \cite{Bergman:2007pm,Dhar:2007bz}, see also~\cite{Casero:2007ae}.},
the gluons deconfine but the chiral symmetry remains
broken. Finally, in the high-temperature phase, the chiral symmetry is
restored. The behavior of low-spin mesons, which we will focus on in
this paper, is described by small fluctuations of the flavor
D8-branes, which can be treated (for $N_f \ll N_c$) as probes embedded in the D4-brane
background. The meson spectra at
zero chemical potential and non-zero temperature have been analyzed
in~\cite{Peeters:2006iu}.

Since the model has a vanishing quark mass, the pions are massless,
and upon turning on an isospin chemical potential the charged pions
immediately want to condense. In the absence of a pion mass, the pion
condensate is stabilized by the non-linear interactions of the chiral
Lagrangian; for any non-zero value of the isospin chemical potential,
the pion goes to its maximal possible value, given that it is a
Nambu-Goldstone boson which lives on a compact target space.  The pion
condensate leads to a small technical problem in our analysis; since
the pion condensate in the Sakai-Sugimoto model corresponds to the
holonomy of the D8-brane gauge field in the holographic $z$ direction, one has
to deal with solutions with~$A_z\not=0$, while usually it is more
convenient to analyze this model in the $A_z=0$ gauge. To avoid this
complication, one can perform an $SU(N_f)_L\times SU(N_f)_R$ global
symmetry transformation (which is a ``large gauge transformation'' in
the Sakai-Sugimoto model) which eliminates the pion vacuum expectation
value, and then continue working in the $A_z =0$ gauge.  This global
symmetry transformation maps a vectorial isospin chemical potential
with a pion condensate into an axial isospin chemical potential
with no pion condensate.  Thus, instead of studying the
effects of the vector-like isospin chemical potential in the presence
of the pion condensate, we can equivalently study the effects of an
axial isospin chemical potential in the trivial $A_z=0$ vacuum. Using
this method, we study the effects of the chemical potential as $\mu_I$
is increased. In particular, we determine the behavior of the masses
of the low-spin mesons as a function of~$\mu_I$.  Note that axial
flavor chemical potentials may be interesting in their own right for
studying the full phase diagram of QCD; as mentioned above, an axial
isospin chemical potential is equivalent to a vector-like isospin
chemical potential by a global symmetry transformation, but one can
also study an axial $U(1)$ chemical potential, and we will describe
how the Sakai-Sugimoto model reacts to such a chemical potential as
well.
\medskip

Our results for the Sakai-Sugimoto model at low temperatures (below the
deconfinement temperature) and
finite isospin chemical potential (in an $SU(2)$ subgroup of the
diagonal $SU(N_f)$ subgroup of the flavor group) are the following. For
small chemical potentials we correctly reproduce the results of the
chiral Lagrangian, and the charged pion condenses to its maximal
value. The isospin chemical potential $\mu_I$ explicitly breaks the isospin symmetry to
$U(1)$; the pion condensate then breaks this $U(1)$ symmetry, but in
the vacuum with the pion condensate there is a new $U(1)$ symmetry
which is an axial subgroup of $U(N_f)_L\times U(N_f)_R$ which is
unbroken. In this phase the isospin charge density is exactly linear
in $\mu_I$.  The interesting physics starts to appear when the
chemical potential becomes of the order of the $\rho$~meson mass. In
this paper we analyze this physics for the special value of the parameters $L
= \pi R$, where the D8-branes are anti-podal to the anti-D8-branes. We
find that there is a critical value of $\mu_I = \mu_{\text{crit}} \approx
1.7\, m_{\rho}$ where the~$\rho$ meson becomes unstable and
condenses.  This condensate breaks the rotational invariance from
$SO(3)$ to $SO(2)$, and also breaks the remaining $U(1)$ symmetry
completely.  We numerically construct the new ground state, and find
that for $\mu_I$ slightly larger than $\mu_{\text{crit}}$, the
condensate behaves as~$\langle \rho \rangle \propto \sqrt{\mu_I-\mu_{\text{crit}}}$,
as expected for a second order phase transition. On
top of the $\rho$ meson condensate, the pion condensate persists as
well.  The analysis of larger values of $\mu_I$, as well as of other
values of $L/R$ (where there is also a non-trivial intermediate
temperature phase
\footnote{The analysis of isospin chemical potentials in the
  high-temperature phase of the Sakai-Sugimoto model is a
  straightforward generalization of the analysis
  of~\cite{Horigome:2006xu,Parnachev:2006ev}; in this phase the
  dynamics involves a condensation of deconfined quarks, and there is
  no difference between the isospin and baryon chemical potentials, as
  well as between vector-like and axial-like chemical potentials. Isospin
  chemical potentials in the intermediate and high temperature phases of
  the Sakai-Sugimoto model were recently analyzed in~\cite{Parnachev:2007bc},
  but the results there do not agree with ours.}),
is postponed to future work. In particular, it would be interesting to
check the stability of the new state we find, and to
see if additional mesons also condense as $\mu_I$ is further
increased. In the Sakai-Sugimoto model (as in all weakly curved
holographic models) the mesons of spin larger than one are much
heavier (by a factor of $\sqrt{\lambda_5/R}$) than the low-spin
mesons; however, when $\mu_I$ reaches the scale of the mass of these
mesons, they may also want to condense. Our analysis is performed in
the approximation of $\mu_I \ll \lambda_5 / R^2$; for larger values of
$\mu_I$ one has to include also DBI and Chern-Simons corrections to the
D8-brane effective action, and it would be interesting to see if these wash out
the rho meson condensate that we found or not.

Let us end with a few remarks concerning the relevance of our analysis
for real (large $N_c$) QCD. The
large isospin chemical potential analysis in QCD reviewed above, and the
subsequent conjecture about the absence of phase
transitions~\cite{Son:2000by}, crucially depends on the underlying
asymptotic freedom of the theory. The Sakai-Sugimoto model, however,
does not exhibit asymptotic freedom (except in its QCD limit), so
apriori it is not clear how this model behaves even for large
$\mu_I$. We do not know if the phase transition described in the
previous paragraph persists when we go to the QCD limit of small
$\lambda_5/R$ or not. However, it is
plausible~\cite{Voskresensky:1997ub,Lenaghan:2001sd,Sannino:2002wp} that a similar phase
transition could occur also in QCD in the regime of
intermediate~$\mu_I$ (of order the $\rho$ meson mass), where the
analysis of~\cite{Son:2000by} does not apply. It would be interesting
to compute corrections to our analysis in inverse powers of
$\lambda_5/R$ (related to $\alpha'$ corrections), to see if these tend to
strengthen or weaken the rho meson condensate, as we go in the direction
of approaching QCD.

We begin in section 2 with a review of the Sakai-Sugimoto model,
focussing on the realization of chiral symmetry and on the corresponding
moduli space of vacua. In section 3 we analyze the isotropic solutions
with chemical potentials, and compare them (for small chemical potentials)
to the expectations from the chiral Lagrangian. In section 4 we analyze
the stability of the isotropic solutions, and find that there is a critical
value of the chemical potential where they become unstable; above that
value we construct a new branch of solutions in which the rotational
symmetry is spontaneously broken. An appendix contains a discussion of some technical
issues in the numerical analysis.

\section{The configurations with no chemical potential}
\subsection{Review of the phases of the Sakai-Sugimoto model}

The Sakai-Sugimoto model \cite{Sakai:2004cn} is the only known example
of a holographic dual for a $3+1$ dimensional gauge theory with a
continuous $SU(N_f)_L\times SU(N_f)_R$ chiral symmetry which is
spontaneously broken. It is derived by considering the decoupling
limit of $N_c$ D4-branes, compactified on a circle of radius $R$ ($x_4
\equiv x_4 + 2 \pi R$) with anti-periodic boundary conditions for the
fermions~\cite{Witten:1998zw}, intersecting with $N_f$ D8-branes at
$x_4=0$ and $N_f$ anti-D8-branes at $x_4=L$. On the field theory side
this gives a $4+1$ dimensional $SU(N_c)$ maximally supersymmetric
gauge theory (with a specific UV completion which will not be relevant
for us), compactified on a circle with anti-periodic boundary
conditions for the adjoint fermions, and coupled to $N_f$ left-handed
fermions in the fundamental representation of $SU(N_c)$ localized at
$x_4=0$, and to $N_f$ right-handed fermions in the fundamental
representation localized at $x_4=L$. In this section we will briefly
review the holographic description of this model (see
\cite{Sakai:2004cn,Aharony:2006da} for more details).

The holographic dual description of this model involves (in the
large $N_c$ limit with fixed $N_f$) $N_f$ probe D8-branes in a
closed type IIA string background with a metric, Ramond-Ramond 4-form
and string coupling given by
\begin{equation}
\label{closedback}
\begin{aligned}
{\rm d}s^2 &= \left(\frac{u}{R_{D4}}\right)^{3/2} \left[-{\rm d}t^2 + \delta_{ij}
{\rm d}x^i {\rm d}x^j + f(u) {\rm d}x_4^2\right] + \left(\frac{R_{D4}}{u}\right)^{3/2}
\left[\frac{{\rm d}u^2}{f(u)}+u^2
{\rm d}\Omega_4^2\right]\,,\\
F_{(4)} &= \frac{2 \pi N_c}{V_4} \epsilon_4,\qquad e^{\phi} = g_s
\left(\frac{u}{R_{D4}}\right)^{3/4},\qquad R_{D4}^3 \equiv \pi g_s
N_c \, l_s^3, \qquad f(u) \equiv 1 -
\left(\frac{\Ukk}{u}\right)^3\,,
\end{aligned}
\end{equation}
where ${\rm d}\Omega_4^2$ is the metric of a unit four-sphere and
$\epsilon_4$ is its volume form, and $g_s$ is related to the $4+1$
dimensional gauge coupling by $g_5^2 = (2\pi)^2 g_s l_s$. The
submanifold spanned by $x_4$ and $u$ has the topology of a cigar
with $u \geq \Ukk$, and requiring that this has a non-singular
geometry gives a relation between $\Ukk$ and $R$,
\begin{equation}
R = \frac{2}{3} \left(\frac{R_{D4}^3}{\Ukk}\right)^{1/2}.
\end{equation}
The background is weakly curved and can be well-approximated by
gravity whenever the 't Hooft coupling $\lambda_5 = g_5^2 N_c$ of
the $4+1$ dimensional gauge theory is much larger than $R$. In the
rest of this paper we will work in this regime. Note that in this
regime the mass gap is of the same order as the scale $1/R$ of the
Kaluza-Klein modes in the gauge theory, so the $3+1$ dimensional
theory does not decouple from the higher Kaluza-Klein modes (as it
does in the opposite limit of $\lambda_5 \ll R$, which is equivalent
to large $N_c$ QCD at energies much smaller than $1/R$). In this
regime the confining string tension $T_s \sim \lambda_5 / R^3$ is
much larger than the scale of the mass gap, enabling an approximation
in which the higher string excitations are ignored.

The probe D8-branes fill the $3+1$ dimensional space-time and the
four-sphere, and trace a curve $u(x_4)$ in the remaining two
coordinates. This curve is a solution to the equations of motion
of the probe D8-branes, with the boundary condition that $u\to
\infty$ at $x_4=0$ (with one orientation of the D8-branes) and at
$x_4=L$ (with the opposite orientation). At $x_4=L/2$ the solution
reaches its minimal value of $u$, which we denote by $u=u_0$. In
the special case of $L = \pi R$, $u_0=\Ukk$ and there is a
simple solution in which $u$ is independent of $x_4$, with
a brane at $x_4=0$ joining smoothly (at $u=u_\Lambda$) with an
anti-brane at $x_4 = \pi R$. In other cases the solution cannot be
written down explicitly, but it satisfies the first order
differential equation
\begin{equation}
\frac{u^4 f(u)}{\displaystyle\sqrt{f(u) + \left(\frac{R_{D4}}{u}\right)^3
\frac{u'^2}{f(u)}}} = u_0^4 \sqrt{f(u_0)}. \end{equation}
As we will review below, the fact that the D8-branes and
anti-D8-branes join together smoothly is a reflection of the spontaneous
breaking of the (classical) $U(N_f)_L\times U(N_f)_R$ chiral
symmetry to the diagonal $U(N_f)_V$.

\begin{figure}[t]
\begin{center}
\includegraphics[width=\textwidth]{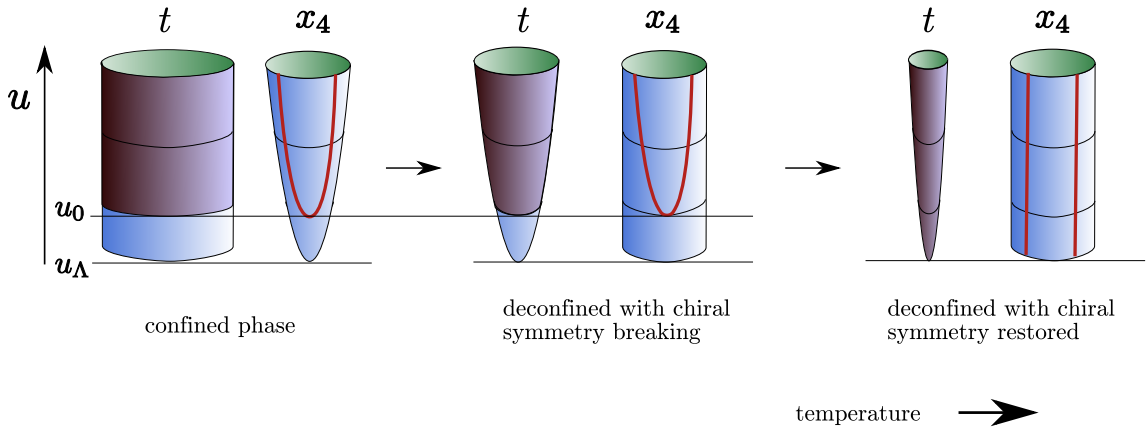}
\end{center}
\caption{The topologies of the background and of the D8-branes in the
  three phases of the Sakai-Sugimoto model. For $L > 0.97 R$, the
  model jumps directly from the first phase to the third
  phase at $T=1/2\pi R$.\label{phases}}
\end{figure}

At finite temperature $T$ we can describe the theory by a solution
of the equations of motion in Euclidean space. There are two
possibilities for the closed string background. For $T < 1 / 2 \pi
R$ the dominant solution is simply the Euclidean continuation of
(\ref{closedback}), with the time direction periodically
identified $t \equiv t + 1 / T$. In this solution the $x_4$ circle
shrinks to zero size at $u=u_\Lambda$ while the Euclidean time circle
never shrinks (see the left-hand side of figure~\ref{phases}). 
For $T > 1 / 2 \pi R$ the solution is the same but
with the roles of $t$ and $x_4$ interchanged, such that the $f(u)$
factor in the metric sits in front of ${\rm d}t^2$. In this phase
$\Ukk$ is related to the periodicity in the Euclidean time
direction,
\begin{equation}
\frac{1}{T} =
\frac{4\pi}{3}\left(\frac{R_{D4}^3}{\Ukk}\right)^{1/2}.
\end{equation}

In the low-temperature phase $T < 1 / 2 \pi R$ the profile of the
D8-branes remains exactly the same as at zero temperature. At
higher temperatures there are two possible profiles for the
D8-branes \cite{Aharony:2006da,Parnachev:2006dn}. For $1 / 2 \pi R
< T < 0.154/L$ (the ``intermediate temperature phase''), the
dominant profile is similar to the one above, and the branes join
together and spontaneously break the chiral symmetry (see the
middle of figure \ref{phases}). The function
$u(x_4)$ obeys a very similar equation to the one above,
\begin{equation}
\frac{u^4 \sqrt{f(u)}}{\displaystyle\sqrt{1 + \left(\frac{R_{D4}}{u}\right)^3
\frac{u'^2}{f(u)}}} = u_0^4 \sqrt{f(u_0)}. \end{equation}
On the other hand, for $T > 0.154/L$ and $T > 1 / 2 \pi R$ (the
``high temperature phase''), the dominant configuration is that of
separate branes at $x_4=0$ and anti-branes at $x_4=L$, which
smoothly end at the horizon $u=\Ukk$ (see the right-hand side
of figure \ref{phases}). In this phase the chiral
symmetry is restored.

\subsection{The chiral symmetry and the moduli space}
\label{s:moduli_space}

In holographic dualities (generalizing the AdS/CFT
correspondence~\cite{mald2}) there is a one-to-one mapping between
gauge symmetries in the bulk and global symmetries of the dual field
theory. More precisely, any gauge field $A_{\mu}(x,u)$ (where $x$ are
the coordinates of the dual field theory, $u$ is the radial direction
and we assume that we have performed a Kaluza-Klein reduction over any
additional compact dimensions) that is present near the boundary $u\to
\infty$ corresponds to a generator of the global symmetry of the
theory. The theory is invariant under gauge transformations
\begin{equation} \label{gaugesymm}
A_{\mu} \to g A_{\mu} g^{-1} + i g \del_{\mu} g^{-1}
\end{equation}
with a gauge parameter $g(x,u)$ (an element of the gauge group $G$)
which goes to the identity element $g=I$ at the boundary. 
``Large gauge transformations''
in which $g(x,u\to \infty) = g_0$ correspond to global symmetry
transformations with parameter $g_0$\footnote{If the $u\to \infty$
limit of $g(x,u)$ gives a result which depends on $x$, then the
transformation is not a symmetry at all.}. If the vacuum is
invariant under such a ``large gauge transformation'' then the
global symmetry is unbroken, and otherwise it is broken.

This correspondence applies both to gauge fields coming from closed
strings and to gauge fields coming from open strings. In the D8-brane
configurations we discussed above, there are two independent $U(N_f)$
gauge fields living near the boundary of space-time; one gauge field
($A_{\mu}^L$) living at $x_4=0$ and another ($A_{\mu}^R$) living at
$x_4=L$. Thus, these theories have a $U(N_f)_L\times U(N_f)_R$ global
symmetry\footnote{The axial $U(1)$ symmetry is actually anomalous so
  it is not a global symmetry. However, in this paper we will only
  discuss the classical theory in the bulk (the leading order in
  $1/N_c$) so this will not be relevant for
  us~\cite{Witten:1979vv,Veneziano:1979ec,Sakai:2004cn,Bergman:2006xn}.\label{fn:anomaly}}.
Note that in this case the gauge fields near the boundary are only
functions of $3+1$ out of the $4+1$ dimensions of the gauge theory,
reflecting the fact that the global symmetry acts on fields which are
localized in $3+1$ dimensions.

In a configuration where the D8-branes are separate from the
anti-D8-branes, as we have in the ``high temperature phase'', we
can perform separate constant gauge transformations (of the form
(\ref{gaugesymm}) with $g=g_0 \in U(N_f)$) for the fields
$A_{\mu}^L$ and $A_{\mu}^R$, and the theory is invariant under
these transformations. Thus, in this phase the full global
symmetry is unbroken.

On the other hand, in a phase where the D8-branes and anti-D8-branes
are connected, the fields $A_{\mu}^L$ and $A_{\mu}^R$ are limits of a
single gauge field living on the D8-branes, and we cannot perform
independent gauge transformations on each of them separately.  We can
still perform a constant gauge transformation $g=g_0$ on the full
gauge field $A_{\mu}(\vec{x},z)$ (here~$z$ is a single-valued
coordinate on the D8-branes, such that $z\to -\infty$ as the D8-branes
approach the boundary at $x_4 = 0$, and $z\to \infty$ as the D8-branes
approaches the boundary at $x_4 = L$; see~\S\ref{s:gfsols} for
details). Near the boundary this acts on both $A_{\mu}^L$ and
$A_{\mu}^R$ by the same gauge transformation $g_0$, so it corresponds
to the vector-like global symmetry in the theory. If we start from a
configuration of the D8-branes which has a vanishing gauge field
$A_{\mu}=0$, then this transformation leaves this configuration
invariant, meaning that the vector-like symmetry is unbroken.

However, suppose that we want to act with a non-vector-like global
symmetry transformation, in which the transformation on
$A_{\mu}^L$ goes to $g_L \in U(N_f)$ near the boundary, while that
of $A_{\mu}^R$ goes to $g_R \in U(N_f)$ (with $g_R \neq g_L$). We
cannot achieve such a transformation with a gauge parameter
$g(x,z)$ which is a constant. The simplest possibility is to
choose $g=g(z)$, such that $g$ is a smooth function with
$g(-\infty)=g_L$ and $g(\infty)=g_R$. However, such a gauge
transformation does not leave our configuration invariant, since
it generates $A_z = i g \del_z g^{-1}$ which is non-vanishing. In
particular, the gauge-invariant holonomy 
\begin{equation}
\label{e:UdefSS}
U = P \exp{(i \int_{-\infty}^{\infty}\!{\rm d}z\, A_z)}\,,
\end{equation}
which was the identity matrix $U = I$ before this transformation,
becomes $U = g_L^{-1} g_R$ after this transformation. Thus, the theory
is not invariant under global symmetry transformations with $g_L \neq
g_R$, reflecting the spontaneous breaking of the chiral symmetry.

Of course, even when the global symmetry is spontaneously broken,
a global symmetry transformation still takes us from one vacuum to
another, equivalent, vacuum of the theory. The theory we are
discussing has a (classical) moduli space of vacua $(U(N_f)\times
U(N_f)) / U(N_f)$, and we can go from any vacuum to any other
vacuum by a global symmetry transformation. Usually one
characterizes the vacua by a matrix $U$ which transforms under the
global symmetry as $U \to g_L^{-1} U g_R$; we see that in the
Sakai-Sugimoto model this matrix is precisely the holonomy
\eqref{e:UdefSS} we discussed above\footnote{Note that in the other vacua the
preserved symmetry is not the vector-like subgroup of
$U(N_f)\times U(N_f)$ but a different subgroup, defined by
$g_L^{-1} U g_R = U$.\label{fn:unbroken}}. The fluctuations in $U$ are usually
written as $U = \exp{(i \pi_a(x) T^a / f_{\pi})}$, where $T^a$ are
the generators of $U(N_f)$ and the ``pions'' $\pi_a(x)$ are the
Nambu-Goldstone bosons; in the Sakai-Sugimoto model these
particles come from fluctuations of the lowest radial mode of
$A_z$ (this statement depends on the gauge; for instance, in the
$U=I$ vacuum one could choose the gauge $A_z=0$ and then the pions
would come from the boundary conditions on the other components of
$A_{\mu}$ instead \cite{Sakai:2004cn,Sakai:2005yt}).

Since all the vacua of the theory are equivalent, it is usually
enough to study the vacuum $U=I$ which is described by the
solution $A_{\mu}=0$. However, in this paper we will be interested
in turning on deformations which explicitly break the global symmetry. Such
deformations destroy the equivalence between the different vacua,
so one should study their effect separately in each vacuum.
Nevertheless, since all the vacua are related by the global
symmetry, we can always study the effect of the deformation on a
vacuum with some non-trivial $U = g_L^{-1} g_R$ by performing the
transformation back to the vacuum $U = I$, but taking into account
the fact that the global symmetry transformation acts also on the
deformation. Thus, instead of studying the effect of the
deformation as a function of the vacuum, we can stay in the vacuum
$U = I$ and study the effect of the deformation as a function of a
global symmetry transformation $(g_L,g_R)$ acting on the
deformation parameters.

\section{Isotropic solutions with flavor chemical potentials}

In this section we will add isospin chemical potentials to the model described
in the previous section. We begin in \S\ref{genchem} with a general
discussion of chemical potentials in the Sakai-Sugimoto model. We continue in
\S\ref{lowene} with an analysis of the effect of chemical potentials
on the chiral Lagrangian, which is the low-energy limit of our
model. In \S\ref{s:gfsols} and \S\ref{s:gfsolslarge} we find isotropic
solutions of the full Sakai-Sugimoto model with chemical potentials,
and we show that they reproduce our expectations from the chiral
Lagrangian. 

\subsection{Flavor chemical potentials in the Sakai-Sugimoto model}
\label{genchem}

As alluded to in the previous section, a gauge field $A_{\mu}$ in
the bulk of a holographic model is dual to a global symmetry
current $J^{\mu}$ in the field theory. As for any other field in
the bulk, the behavior of the gauge field near the boundary
includes independent non-normalizable and normalizable terms; in
the Sakai-Sugimoto model this behavior is given
by (as $u\to \infty$, in the gauge $A_u=0$)
\begin{equation} \label{anubc}
A_{\nu}(x,u) \to B_{\nu}(x) \left(1 + {\cal O}(\frac{1}{u})\right)
+ \rho_{\nu}(x) u^{-3/2} \left(1 + {\cal O}(\frac{1}{u})\right).
\end{equation}
From here on we use conventions with $\mu,\nu=0,1,2,3$, $i,j=1,2,3$.
Again, as for any other field, the coefficient $B_{\nu}(x)$ of the
non-normalizable mode is interpreted as a source term
in the Lagrangian of the field theory of the form $\int\!{\rm d}^4 x\,
B_{\nu}(x) J^{\nu}(x)$, while $\rho_{\nu}(x)$ is proportional to
the vacuum expectation value of $J_{\nu}(x)$. The freedom in
performing gauge transformations in the bulk does not affect the
source term in the Lagrangian due to the conservation equation
$\del_{\nu} J^{\nu}(x) = 0$. Adding a chemical potential to the
field theory corresponds to adding a constant source for $J^0$, so
it corresponds to boundary conditions of the form (\ref{anubc})
with $B_{\nu}(x) = \mu \delta_{\nu,0}$.

As described above, the gauge field living on the D8-branes in the
Sakai-Sugimoto model comes near the boundary at two independent
positions, which we denoted $z\to -\infty$ and $z\to \infty$.
Thus, for this gauge field we have two independent chemical
potentials $A_{\nu}(x,z\to -\infty) = \mu_L \delta_{\nu,0}$ and
$A_{\nu}(x,z\to \infty) = \mu_R \delta_{\nu,0}$, where $\mu_L$ and
$\mu_R$ are matrices in the adjoint of $U(N_f)_L$ and $U(N_f)_R$,
respectively\footnote{We will use conventions in which $A_0$ is
the Lorentzian gauge field, such that $A_0$ and $\mu$ are real.}.

In the high-temperature phase, we have two separate stacks of
D8-branes. The stack at $x_4=0$ only feels $\mu_L$ and reacts to it,
and the stack at $x_4=L$ only feels $\mu_R$ and reacts to it. These
chemical potentials lead to a non-trivial gauge field on the D8-branes
which leads to a specific charge density $\rho_0(x)$ which is linear in
the chemical potential (for small $\mu$), as described
in~\cite{Horigome:2006xu,Parnachev:2006ev}. For small enough $\mu$
(compared to $\lambda_5 T^2$) so that we can ignore the higher order
terms in the DBI action, the solution in this phase has $F_{0u,L} = -
3 \mu_L u_\Lambda^{3/2} / 2 u^{5/2}$ and $F_{0u,R} = - 3 \mu_R
u_\Lambda^{3/2} / 2 u^{5/2}$, leading to a change in the action
\begin{equation} \label{deltashigh}
\Delta S \propto - \tr(\mu_L^2+\mu_R^2) u_\Lambda^{3/2}.
\end{equation}

In this paper we will analyze the lower temperature phases, where
the global symmetry is broken. It is then often convenient to
change the notation from left and right chemical potentials to
vector and axial chemical potentials, defined (in an arbitrary
normalization) by $\mu_L = \mu_V - \mu_A$ and $\mu_R = \mu_V +
\mu_A$. The vector-like chemical potential $\mu_V$ is even under
$z \to -z$, while the axial-like chemical potential $\mu_A$ is
odd. Note that the $U(1)$ components of $\mu_V$ and $\mu_A$ are
the chemical potentials for the baryon number and the axial
$U(1)$, respectively. The non-Abelian components correspond to
isospin-like symmetries, either vector or axial. When studying the
isospin we will usually limit ourselves to an $SU(2)$ subgroup of
$U(N_f)$, and turn on chemical potentials proportional to
$\sigma_3$ in this subgroup; the generalization to arbitrary
chemical potentials (in the Cartan subalgebra of $SU(N_f)$) is
straightforward.

Our definition of $\mu_V$ and $\mu_A$ is natural in the $U=I$ vacuum
where the vector-like subgroup is unbroken
(see footnote~\ref{fn:unbroken}). Under the $U(N_f)\times U(N_f)$
global symmetry, the chemical potentials transform as $\mu_L \to
g_L^{-1} \mu_L g_L$ and $\mu_R \to g_R^{-1} \mu_R g_R$. The $U(1)$
components of the chemical potentials are invariant, but the $SU(N_f)$
components are not. As discussed at the end of \S\ref{s:moduli_space},
we can rotate any vacuum to $U=I$ by an appropriate global symmetry
transformation, but we have to remember to act with the same
transformation also on the chemical potentials. In particular,
studying the same chemical potential in different vacua translates
into studying different chemical potentials in the $U=I$ vacuum.

\subsection{Chemical potentials in the chiral Lagrangian}
\label{lowene}

The spectrum of mesonic states in the Sakai-Sugimoto model at zero
temperature \cite{Sakai:2004cn} includes massless pions (which are the
Nambu-Goldstone bosons of the chiral symmetry breaking) and other
massive mesons. At low enough energies the theory can be described
just by the effective action of the pions; on general grounds this is
given by the chiral Lagrangian, which depends on the matrix $U$ which
we described in the previous section (this was explicitly derived from
the Sakai-Sugimoto model in \cite{Sakai:2004cn})\footnote{At higher
  energies the effective theory of the pions includes also higher
  derivative terms. Including the leading higher correction, which
  comes just from the Yang-Mills action on the D8-branes, leads
  to the Skyrme model with a four-derivative term suppressed by the
  scale $1/L$. Note that this scale is of the same order as the mass
  of the massive mesons, so at this scale one cannot necessarily
  trust an effective action which includes only
  the pions. The full effective action of the
  Sakai-Sugimoto model includes also many additional
  higher derivative terms which are governed by the confining string 
  scale, so at energies of the order of the string scale we can no
  longer use any effective action.}.
  For large
enough temperatures one expects additional degrees of freedom to
become important. However, in the low temperature and intermediate
temperature phases it is still true \cite{Aharony:2006da} that the
only light states which are charged under the flavor symmetry\footnote{
As discussed in the introduction, we will not include the baryon number
symmetry, which behaves very differently from the other flavor
symmetries, in our analysis.} are the
pions, and there is a gap to any other charged states (in the low temperature
phase the other states are just massive mesons, while in the
intermediate temperature phase they include also massive deconfined
quarks). Thus, in these phases that we are interested in, the low
energy dynamics is still captured by the chiral Lagrangian.  In
particular, the response of the theory to small chemical potentials
should also be captured by this Lagrangian. Thus, in this subsection
we analyze the response of the chiral Lagrangian to chemical
potentials. In the next subsection we will verify that the
Sakai-Sugimoto model leads to the same results for small chemical
potentials. Our discussion for vector-like chemical potentials is a
special case of the discussion in~\cite{Splittorff:2000mm} (with
vanishing quark masses), but we will add also axial-like chemical
potentials. As mentioned above (footnote~\ref{fn:anomaly}), since we
are in the large $N_c$ limit we will ignore the anomaly in the axial
$U(1)$.

The chiral Lagrangian may be written as
\begin{equation}\label{lchiral}
{\cal L}_{\text{chiral}} = \frac{f_{\pi}^2}{4} \Tr(D_{\nu} U D^{\nu}
U^{\dagger}).
\end{equation}
The field $U$ is an $N_f\times N_f$ unitary matrix transforming under
the global symmetry as $U\to g_L^{-1} U g_R$. The covariant
derivative $D_{\nu}$ is just the normal derivative in the absence
of chemical potentials. However, in the presence of chemical
potentials it is modified to
\begin{eqnarray}
D_{\nu} U &=& \del_{\nu} U - i \delta_{\nu,0} (\mu_L U - U
\mu_R)~~ = \del_{\nu} U - i \delta_{\nu,0} ([\mu_V, U] -
\{\mu_A, U\}),  \nonumber \\
D_{\nu} U^\dagger &=& \del_{\nu} U^\dagger + i
\delta_{\nu,0} ( U^\dagger \mu_L  - \mu_R U^\dagger) = \del_{\nu}
U^\dagger - i \delta_{\nu,0} ([\mu_V, U^\dagger] +
\{\mu_A, U^\dagger\}).
\end{eqnarray}
Note that the baryon number chemical potential $\mu_V \propto I$
does not appear in the action at all; this is not surprising since
none of the light states in the model carry a baryon number
charge. Thus, the baryon number chemical potential has no effect
on the classical low-energy action. 

The potential energy arising from the Lagrangian (\ref{lchiral})
is
\begin{equation}
V_{\text{chiral}} = 
\frac{f_{\pi}^2}{4} \Tr\left(([\mu_V, U] - \{\mu_A,
U\}) ([\mu_V, U^{\dagger}] + \{\mu_A, U^{\dagger}\})\right).
\end{equation}
The charge density is :
\begin{eqnarray}
\rho_V &=& \frac{\partial{\mathcal L}_{\text{chiral}}}{\partial \mu_{V}}
= -\frac{f_{\pi}^2}{2} (U (\mu_V+\mu_A) U^{\dagger} +
U^{\dagger} (\mu_V-\mu_A) U - 2 \mu_V),\nonumber \\
\rho_A &=& \frac{\partial {\mathcal L}_{\text{chiral}}}{\partial \mu_{A}}
=\frac{f_{\pi}^2}{2} (U (\mu_A+\mu_V) U^{\dagger} + U^{\dagger}
(\mu_A-\mu_V) U + 2 \mu_A).
\end{eqnarray}

We can now analyze the dynamics in various cases. We can start by
having a chemical potential just for the axial $U(1)$, $\mu_A = \mu_a
I / \sqrt{N_f}$. In this case the potential is simply a constant
$V_{\text{chiral}} = -f_{\pi}^2 \mu_a^2$ so there is no potential
generated on the moduli space, and the axial charge density is given
by $\rho_a \equiv \tr(\rho_A) / \sqrt{N_f} = 2 f_{\pi}^2 \mu_a$. The
axial $U(1)$ chemical potential has precisely the same, rather trivial,
effect also when additional chemical potentials are turned on, so we
will assume it is zero from here on.

Next, let us turn on a vector-like isospin chemical potential $\mu_V =
\mu_I \sigma_3 / 2$ \cite{Splittorff:2000mm}. Our potential then takes the
form
\begin{equation}
V_{\text{chiral}} = \frac{f_{\pi}^2 \mu_I^2}{8} \Tr(\sigma_3 U \sigma_3
U^{\dagger} - 1).
\end{equation}
This is maximized by configurations where $U$ is along the
identity and $\sigma_3$ directions, namely $U_{\text{max}} = e^{i\alpha}
(\cos(\beta) I + i \sin(\beta) \sigma_3)$, and minimized by
configurations in the $\sigma_{1,2}$ directions, $U_{\text{min}} =
e^{i\alpha} (\cos(\beta) \sigma_1 + \sin(\beta) \sigma_2)$. Thus,
the minimal energy configurations in this case will involve
matrices of the form $U_{\text{min}}$. This is natural since there are
massless pions charged under the isospin symmetry, and they want
to condense to their maximal possible value 
when we turn on the isospin chemical potential. The
isospin charge density in the $U_{\text{min}}$ configurations is given by
$\tr(\rho_V \sigma_3) = 2 f_{\pi}^2 \mu_I$, while in the $U_{\text{max}}$
configurations it vanishes. Note that in the vacua $U=U_{\text{min}}$ the
$U(1)$ symmetry corresponding to the chemical potential that we turn on
is broken.

It is just as simple to analyze the case where we have both axial
and vector isospin chemical potentials; assuming that both of them
are in the $\sigma_3$ direction ($\mu_V = \mu_I \sigma_3 / 2$ and
$\mu_A = \mu_{A,I} \sigma_3 / 2$), we find a potential
\begin{equation}
V_{\text{chiral}} = \frac{f_{\pi}^2}{8} [(\mu_I^2 - \mu_{A,I}^2)
\Tr\big(\sigma_3 U \sigma_3 U^{\dagger}\big) - (\mu_I^2
+\mu_{A,I}^2) \Tr\big(1\big)] \, .
\end{equation}
Thus, if $\mu_I^2 > \mu_{A,I}^2$ the minimum is still at the same
$U_{\text{min}}$, and the vacuum energy turns out to be independent of
$\mu_{A,I}$, while if $\mu_I^2 < \mu_{A,I}^2$ then the minimum is
at one of the $U_{\max}$ configurations, and the vacuum energy is
independent of $\mu_I$. In all of these cases, the vector-like
charge density vanishes in the $U=I$ vacuum, while the axial-like
charge density is $\tr(\rho_A \sigma_3) = 2 f_{\pi}^2 \mu_{A,I}$.

\begin{figure}[t]
\begin{center}
\includegraphics[width=.9\textwidth]{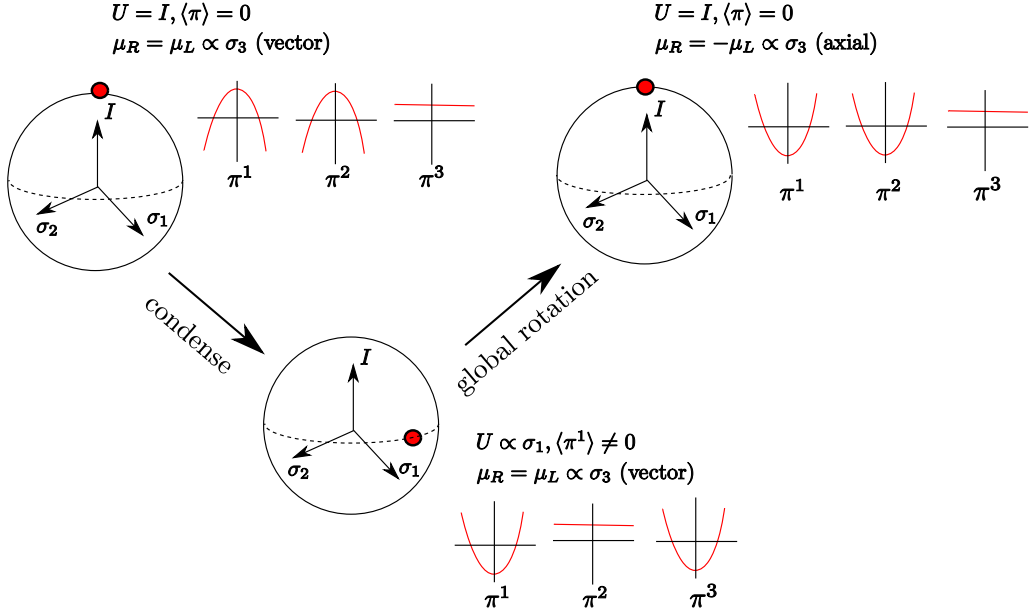}
\end{center}
\caption{Introducing a vectorial isospin chemical potential leads to an
  instability for two of the pions (top left), and the theory wants to
  sit at a new minimum~$U=U_{\text{min}}$
  (bottom). A global symmetry transformation brings us back to the vacuum~$U=I$; this
  rotation changes the vector chemical potential to an axial one (top right).}
\label{different_vacua}
\end{figure}

An important point is that, as we mentioned at the end of
\S\ref{s:moduli_space}, we can study all these vacua also by
transforming them to the trivial vacuum $U=I$. Suppose we wish to
study the theory with a purely vector-like isospin chemical potential, $\mu_L =
\mu_R = \mu_I \sigma_3 / 2$. As discussed above, in this case $U=I$ is a
maximum of the scalar potential, so it corresponds to an unstable
solution to the equations of motion. The stable solutions are at
$U=U_{\text{min}}$.  We can transform this to $U=I$ in various
ways. For example, we can choose $g_L = U_{\text{min}}$, $g_R = I$;
this transforms the vacuum to $U=I$, and the chemical potentials to a
purely axial form $\mu_L = -\mu_R = -\mu_I \sigma_3 / 2$. Other choices give
similar results (related by the $U(N_f)_V$ symmetry which is unbroken
in the $U=I$ vacuum).  Thus, we can study the behavior at the minima
of the scalar potential with a vector-like isospin chemical potential
by analyzing the vacuum $U=I$ with a purely axial isospin chemical
potential, see figure \ref{different_vacua}.  
Note that this is consistent with the fact that $U=I$ is a
minimum of the potential in the latter case.\footnote{Note also that if we
transform a general value of $U$ (with a vector-like isospin chemical
potential), which is neither of the form $U_{\text{min}}$ nor of the
form $U_{\text{max}}$, to $U=I$, we would get vector-like and
axial-like chemical potentials that do not commute with each other;
for such chemical potentials $U=I$ is not an extremum of the scalar
potential, so it is not a static solution to the equations of motion.}
Note also that even though the pion condensate breaks the original
$U(1)$ symmetry preserved by a purely vector-like isospin chemical
potential, there is a new $U(1)$ global symmetry which is unbroken
in each of the vacua $U=U_{\text{min}}$; after the global symmetry
transformation to $U=I$ this is simply the vector-like $U(1)$ subgroup
of the isospin symmetry (but in the original variables it is an
axial-like symmetry).

\subsection{Isotropic solutions for small chemical potentials}
\label{s:gfsols}

In this section we will look for solutions to the equations of
motion of the Sakai-Sugimoto model in the presence of the various
chemical potentials discussed above. As discussed in \S\ref{genchem}, such
solutions involve configurations in which $A_0$ is non-vanishing. In
this section we will only look for solutions which are
translationally and rotationally invariant. Such solutions depend
only on the time $t$ and on the radial coordinate $z$, and must have
$A_i=0$.

We will start with the solutions for small $\mu$, which can be
compared to those of the chiral Lagrangian, and in
\S\ref{s:gfsolslarge} we will generalize this to larger values of
$\mu$. For small $\mu$ we can approximate the action for the gauge
fields on the D8-brane by the Yang-Mills action; from the form of the
solutions that we will find, it is easy to see that the DBI corrections to
the YM action start becoming important when $\mu \sim \lambda_5 /
L^2$, so our solutions in this subsection are valid for $\mu \ll
\lambda_5 / L^2$. Moreover, we will only work to the leading
non-trivial order in $\mu$, so we will ignore the backreaction of the
gauge fields on the brane embedding $u(x_4)$ which is of order $\mu^2$
(this is not present when $L = \pi R$ but is present in all other
cases).  Since we are interested in solutions with $A_i=0$, the CS
term will play no role, so we will ignore it in this section.

As mentioned above, it is useful to choose a coordinate $z$ which is
single-valued along the D8-branes; we will choose to define such a
coordinate by
\begin{equation}
u = (u_0^3 + u_0 z^2)^{1/3}\,,\qquad \frac{\partial u}{\partial z} =
\frac{2}{3}\frac{u_0 z}{u^2}\,,
\end{equation}
where $z$ (and $\partial u / \partial z$) are positive for $x_4 >
L/2$ and negative for $x_4 < L/2$ ($z=0$ corresponds to the tip of
the branes at $u=u_0$, while  $z=\pm\infty$ correspond to the two
asymptotic regions on the branes). Using this coordinate, we can
write down the YM action density for the gauge fields on the
D8-branes in the low temperature phase as
\begin{equation}
S^{\text{low}}_{\text{D8}} =
  \tilde{T}\! \int_{-\infty}^{\infty}\!{\rm d}z\,
     \bigg[ \frac{R_{D4}^3 |z|}{4 u^{5/2}}\tr(F_{\mu\nu}^2) \sqrt{\gamma}
           + \frac{9}{8}\frac{u^{9/2}}{|z| u_0^2} \tr(F_{\mu z}^2) \frac{1}{\sqrt{\gamma}}\bigg]
\end{equation}
with~$\tilde{T} = \frac{8}{3} {\hat T}_8 R_{D4}^{3/2} u_0 (2\pi
\alpha')^2 / g_s$ (where, as in \cite{Aharony:2006da}, ${\hat T}_8$
includes the tension of the D8-brane and the volume of the
four-sphere), and
\begin{equation}
\gamma(u) \equiv \frac{u^8}{u^8 f(u) - u_0^8 f(u_0)}\,.
\end{equation}
In the intermediate temperature phase, where the gluons are
deconfined but the chiral symmetry is still broken, the action
density reads
\begin{equation}
S^{\text{int}}_{\text{D8}} =
  \tilde{T} \int_{-\infty}^{\infty}\!\!\!\!\!{\rm d}z\, \sqrt{f}\,
     \bigg[ \frac{R_{D4}^3 |z|}{4 u^{5/2}} \left\{ \tr(F_{ij}^2) + \frac{2}{f}
  \tr(F_{0i}^2) \right\} \sqrt{\gamma}  + \frac{9}{8}\frac{u^{9/2}}{|z| u_0^2} \left\{ \frac{1}{f} \tr(F_{0 z}^2) +
  \tr(F_{iz}^2) \right\} \frac{1}{\sqrt{\gamma}}\bigg]\,.
\end{equation}
Note that in the limit~$u_0=\Ukk$ these expressions simplify because
then $\gamma = f^{-1}$ and $f = u_0 z^2 / u^3$.
These actions assume that the brane configuration is the same as it
was in the absence of the chemical potential; as mentioned above, we
expect the chemical potential to generate gauge fields of order
$\mu$, which should back-react on the configuration of the brane at
order $\mu^2$.

As discussed above, we are looking for solutions with $A_i=0$, so
the relevant part of the action is
given by the $F_{0 z}^2$ terms, which appear with a measure~$g_2(z)$
given by
\begin{equation}
g_2(z) = \frac{9}{8}\frac{1}{u_0^2} \tilde{T} \times
\begin{cases}
(u_0^3 + u_0\, z^2)^{3/2} \frac{1}{|z| \sqrt{\gamma}}  & \quad \text{low $T$}\,,\\[1ex]
(u_0^3 + u_0\, z^2)^{3/2} \frac{1}{|z| \sqrt{f \gamma}} &
\quad\text{intermediate $T$}\,.
\end{cases}
\end{equation}
The large~$|z|$ behavior is obviously the same for both cases.

For simplicity, we will work in the $A_z=0$ gauge. As discussed
above, this means that we are at a specific point $U=I$ in the
moduli space, but this point is related to all other points by the
global symmetry transformation so we can still discuss the most
general case. Naively, going to $A_z=0$ gauge means that we are setting
the full pion field to zero, and not just its zero modes, but in fact this is
not true; as discussed in \cite{Sakai:2004cn}, in this gauge the
derivatives of the pions show up through the boundary values of
$A_{\mu}$.

Thus, we wish to solve the equations of motion with an ansatz of the
form
\begin{equation}
\label{az1}
\begin{aligned}
&A_z =0 \, , \quad A_i = 0\,,  \\[1ex]
&A_0 = A_0^V (z,t) + A_0^A (z,t) \, , \quad F_{0z} = -\partial_z A_0
\, ,
\end{aligned}
\end{equation}
where we separated the field $A_0$ into its even part (under $z \to
-z$) $A_0^V$ (related to the vector-like symmetry) and its odd part
$A_0^A$ (related to the axial symmetry).  As discussed in
\S\ref{genchem}, for a solution with a given chemical potential, $A_0$
should approach $\mu_L$ as $z \to -\infty$ and $\mu_R$ as $z \to
\infty$, and the coefficient of the normalizable mode of $A_0$ at
infinity (which we called $\rho_0(x)$ in (\ref{anubc})), which is
proportional to $1/z$, is proportional to the charge density.

With the ansatz~\eqref{az1}, the equations of motion become
\begin{equation}
\label{eome}
\partial_0 \partial_z A_0 - i [A_0, \partial_z A_0] = 0 \,, \quad
\partial_z \big(g_2(z)\, \partial_z A_0\big) = 0 \, .
\end{equation}
The second equation implies that the general solution for the field
strength takes the form
\begin{equation}
\label{SOLU} F_{0z} = \frac{\tilde{f}_A(t)}{g_2(z)} \, \quad
\Rightarrow \quad A_0 = f_A(t) \int_0^z \frac{{\rm d}
\tilde{z}}{g_2(\tilde{z})} \left(\int_0^\infty \frac{{\rm d}
\tilde{z}'}{g_2(\tilde{z}')}\right)^{-1} + f_V(t) \, ,
\end{equation}
for some $U(N_f)$-valued functions $f_V(t)$ and $f_A(t)$, which must
obey the additional equation of motion
\begin{equation}\label{ffeom}
\partial_0 f_A(t) - i [f_V(t), f_A(t)] = 0.
\end{equation}
Note that the contribution of $f_V$ to $A_0$ is even under $z\to -z$
(in fact it is independent of $z$) so it includes any vector-like
chemical potential and charge density, while the contribution of
$f_A$ is odd and includes the axial chemical potential and charge
densities. In the special case of the low temperature phase with $L
= \pi R$ such that $u_0 = u_\Lambda$, it is easy to perform the
integral in (\ref{SOLU}) explicitly, and to write the general
solution in the form $A_0 = 2 f_A(t) \arctan(z/u_\Lambda) / \pi + f_V(t)$ (where
one still has to impose (\ref{ffeom})).

Near the boundary, the solution (\ref{SOLU}) approaches $f_V(t) \pm
f_A(t)$ at $z=\pm \infty$. In the case that the solution is
time-independent, we thus identify $f_V = \mu_V$ and $f_A = \mu_A$,
and we find such a solution whenever $[\mu_V,\mu_A] = 0$. This agrees
with the expectation described in \S\ref{lowene}, that for commuting
$\mu_V$ and $\mu_A$, $U=I$ should be a static (not necessarily
stable) solution. By examining
the terms going as $1/z$ near the boundary, we see that these
solutions have $\rho_V = 0$ and $\rho_A = C \mu_A$ for some constant
$C$ that we will compute below. Again, this precisely agrees with the
expectations described in \S\ref{lowene}. Note that in the purely
vector-like case the field strength vanishes identically.

The conclusions of the previous paragraph hold both for the $U(1)$
and for the $SU(N_f)$ components; as in the chiral model, the $U(1)$
components of the gauge field are independent of the $SU(N_f)$
components so we can discuss them separately. In the $SU(N_f)$ case,
if we take $\mu_V = \mu_I \sigma_3/2$ and $\mu_A = \mu_{A,I}
\sigma_3/2$, then, as discussed in~\S\ref{lowene}, the solution we find is either
a maximum or a minimum, depending on the sign of $\mu_I^2-\mu_{A,I}^2$.
However, this does not affect the form of the solution, but just the
mass spectrum of small fluctuations around the solution.

As we also discussed in~\S\ref{lowene}, when $[\mu_V, \mu_A] \neq 0$ we do not expect
to find a static solution, and indeed we see from (\ref{ffeom}) that
there is no such solution, but that the time derivative of $f_A$
must be non-zero (even as $z \to \pm \infty$). This implies that the
time-derivative of the pion field must be non-zero in this case,
consistent with the fact that we are not sitting at an extremum of
the pion potential. It is easy to write simple solutions to
(\ref{ffeom}) which describe oscillations around the minima of the
pion potential, but we expect that when we include interactions (at
order $1/N_c$) such solutions will eventually decay to the minima,
so they are not very interesting and we will not discuss them
further.

In order to compute the precise coefficient $C$ appearing
in our expression above for the charge density, we need to compute
the derivative of the action of our configurations with respect to
$\mu_A$ (it is obvious that the action is independent of $\mu_V$).
By plugging (\ref{SOLU}) into the action density, we obtain
\begin{equation}
S = 2 \Tr(\mu_A^2) \left(\int_0^{\infty}
\frac{{\rm d}\tilde{z}}{g_2(\tilde z)}\right)^{-1},
\end{equation}
so that the axial charge density is given by
\begin{equation}
\rho_A = 4 \mu_A \left(\int_0^{\infty} \frac{{\rm d}\tilde{z}}{g_2(\tilde
z)}\right)^{-1}.
\end{equation}
Comparing this with our expressions of the previous section, we
obtain
\begin{equation}
f_{\pi}^2 = 2 \left(\int_0^{\infty} \frac{{\rm d}\tilde{z}}{g_2(\tilde
z)}\right)^{-1}.
\end{equation}
This is precisely the same expression that one obtains by computing the
kinetic term for the holonomy matrix $U$ (defined in~\eqref{e:UdefSS})
in the Sakai-Sugimoto model~\cite{Sakai:2004cn}; note that a
configuration with varying holonomy $U$ is described in this model by
\begin{equation} F_{\mu z} = - \frac{i}{g_2(z)} U^{-1} \partial_{\mu}
U \left(\int_{-\infty}^{\infty} \frac{{\rm d}\tilde{z}}{g_2(\tilde
z)}\right)^{-1}. \end{equation}
For $L = \pi R$ we have \cite{Sakai:2004cn} $f_{\pi}^2 \propto
\lambda_5 N_c / R^3$. For $L \ll R$, on the other hand, the scale
$R$ drops out of the analysis and we have $f_{\pi}^2 \propto
\lambda_5 N_c / L^3$. Since it is not clear how to directly compute
the chiral condensate in the Sakai-Sugimoto model, $f_{\pi}$ is the
best estimate that we have for the scale of chiral symmetry breaking
in this model.

We can now use the results above to analyze the response of the
theory to various chemical potentials. In the case of a baryon
number chemical potential, we see that its only effect is a shift in
the constant mode of $A_0$, which has no effect on the physics. This
is as we expect (for small $\mu$), since we did not include in our analysis any
fields which are charged under the baryon number symmetry. For
vector-like isospin chemical potentials we saw that we have similar
solutions with constant values of $A_0$; these solutions have no
field strength, but they affect the theory through the commutator
terms in the action. As discussed above, these solutions (with
$U=I$) should be unstable maxima of the pion potential, and the
spectrum of fluctuations around these solutions contains tachyonic
modes. If we want
to analyze the minima $U_{\text{min}}$ of the pion potential with a vector-like
isospin chemical potential then, as discussed above, we can
transform them into the $U=I$ configuration with an axial-like
chemical potential. These configurations describe a charged pion 
which condensed to its maximal value (it is stabilized by the
non-linear interactions in the chiral Lagrangian, associated
with the finite moduli space of the pions). 
In these configurations there is a
non-trivial field strength, leading both to a non-trivial free
energy density and to a non-trivial charge density, coming from the
condensation of the charged pions. These configurations are expected
to be stable, at least for small chemical potentials where the
truncation to the chiral Lagrangian is a good approximation; we
will analyze their stability at chemical potentials of the order of
the mass gap in section \ref{s:stability}. 

\subsection{Isotropic solutions for large flavor chemical potentials}
\label{s:gfsolslarge}

The solution~\eqref{SOLU} for axial chemical potentials
found in the previous subsection is valid only for small~$\mu$, or more
precisely, for~$\mu \ll \lambda_5/L^2$. Our analysis ignored both the DBI
corrections to the Yang-Mills action, and the fact that the
back-reaction of the gauge fields which are generated will change
the brane embedding. In this subsection we will generalize this
analysis to larger values of $\mu$. In general,
the effective action on the D8-brane includes also additional
corrections beyond the DBI action, involving derivatives of the
gauge fields. In the configurations we will be discussing in this
paper, the corrections coming from derivatives in the radial
direction are small as long as the gravity approximation is valid,
namely for $\lambda_5 \gg R$. Corrections coming from derivatives
in space-time directions will be small as long as these derivatives
are smaller than the string scale $\sim \lambda_5 / R^3$; at higher
energies or momenta than this we must use the full string theory and
we cannot use any effective action. However, for the static solutions
described in this section, the DBI action is valid for arbitrarily
large values of $\mu$.

We will solve
the equations of motion here for a single D8-brane; as long as the chemical
potential is a diagonal matrix, the full solutions (in the $U=I$
vacuum) will be diagonal, so they will just be $N_f$ copies of the
single brane solution.

In the low temperature phase, the relevant terms in the action for a
single D8-brane are
\begin{equation}
\label{ST=0}
S = \frac{\hat{T}_8}{g_s} \int_0^L \!{\rm d}x_4\, u^4\, \sqrt{f(u)}\,
     \sqrt{1 + \left( \frac{R_{\text{D4}}}{u}\right)^{3}
     \frac{u'{}^2 - f(u) (F_{04})^2}{f(u)^2} } \, .
\end{equation}
For simplicity we will work here in the $u$ coordinate instead of in
the $z$ coordinate that we used before; the translation between the
two coordinate systems is straightforward.  One equation of motion
can be obtained from the ``Hamiltonian'' associated to translations
in~$x_4$,
\begin{equation}
\label{HaT0}
H = \frac{\displaystyle  u^4\,\sqrt{f(u)} }{\displaystyle
 \sqrt{1 + \left( \frac{R_{\text{D4}}}{u}\right)^{3}
     \frac{u'{}^2 - f(u)\,(F_{04})^2}{f(u)^2} }} \, \, ,
\end{equation}
which is conserved, $\partial_4 H = 0$. In addition, the equation of
motion for the gauge potential $A_0$ implies
\begin{equation}
\label{eomA0T0} E \equiv \frac{\displaystyle u^4\,
F_{04}}{\displaystyle \left(\frac{u}{R_{\text{D4}}}\right)^3
\sqrt{f(u)}
 \sqrt{1 + \left( \frac{R_{\text{D4}}}{u}\right)^{3}
     \frac{u'{}^2 - f(u)\,F_{04}^2}{f(u)^2}}} \, = \text{const}\,.
\end{equation}
Dividing~\eqref{eomA0T0} by~\eqref{HaT0} we get for the electric
field
\begin{equation}
\label{e:EzeroT} F_{04} = \frac{E}{H}
\left(\frac{u}{R_{\text{D4}}}\right)^3  f(u) \, .
\end{equation}
Note that $F_{04}$ is symmetric under $z \to -z$, which means
that $F_{0z}$ is anti-symmetric, so this is an axial-like solution
\footnote{This axial solution was also obtained
  in~\cite{Horigome:2006xu}, but incorrectly interpreted as a
  vector-like
  baryon chemical potential. When the configuration is made symmetric
  under~$z\rightarrow -z$, the resulting gauge field is no longer
  continuous at the tip of the brane~\cite{Parnachev:2006ev}.}.
  Of course, the vector-like solution is still given simply by
  $A_0 = {\rm const}$, $E=0$.

In order to express this in terms of our field theory parameters we
need to find the shape~$u(x_4)$. The brane minimum is by definition
at $u=u_0$, where $u'=0$. This fixes~$H$ in terms of~$u_0$ to be
\begin{equation} \label{forh}
H = \sqrt{ u_0^8 f(u_0) + E^2 \, f(u_0)
  \left(\frac{u_0}{R_{\text{D4}}}\right)^3 }\,.
\end{equation}
We can then solve for~$u'$ from the expression for~$H$, which yields
\begin{equation}
\label{up}
u' = \sqrt{ u^8 f(u) - u_0^8 f(u_0) +
  \frac{E^2}{R_{\text{D4}}^3} \big( f(u) u^3 - f(u_0) u_0^3 \big)}
  \left(\frac{u}{R_{\text{D4}}}\right)^{3/2}\,f(u)\, \frac{1}{H}\,.
\end{equation}

As before, for a vector-like chemical potential the solution is
simply $A_0 = \mu_V$ and $F_{04}=0$. For an axial-like chemical
potential\footnote{We remind the reader that this case is equivalent
to a vector-like isospin chemical potential when the theory is expanded
around the minimum of the pion potential.} 
we have the following relations between the parameters of
the field theory, $\mu_A$ and $L$, and the parameters $E$ and $u_0$
of the solutions found above :
\begin{equation}
\begin{aligned}
\mu_A &= \frac{1}{2} \int_0^L dx_4 \partial_4 A_0 =\cr &= - E
\int_{u_0}^{\infty} du \left(\frac{u}{R_{D4}}\right)^{3/2} \left[
u^8 f(u) - u_0^8 f(u_0) +
  \frac{E^2}{R_{\text{D4}}^3} \big( f(u) u^3 - f(u_0) u_0^3 \big)
  \right]^{-1/2},
\end{aligned}
\end{equation}
\begin{equation}
\begin{aligned}
L &= \int dx_4 = 2 \int_{u_0}^{\infty} \frac{du}{u'} =\cr &= 2 H
\int_{u_0}^{\infty} \frac{du}{f(u)}
\left(\frac{R_{D4}}{u}\right)^{3/2} \left[ u^8 f(u) - u_0^8 f(u_0) +
  \frac{E^2}{R_{\text{D4}}^3} \big( f(u) u^3 - f(u_0) u_0^3 \big)
  \right]^{-1/2},
\end{aligned}
\end{equation}
where in the last equation $H$ should be expressed in terms of $u_0$
using \eqref{forh}. It is easy to see that by expanding these
solutions to leading order in $\mu$ (and in $E$) one reobtains the
solutions we discussed in the previous subsection \eqref{SOLU}. 

Note that if we have generic isospin chemical potentials with 
$N_f > 2$ and the values of $\mu_A^2$ are not
equal along the diagonal, then generally
the D8-branes will no
longer be overlapping but will have different functions $u(x_4)$
except near the boundary. This was not visible in the previous
subsection since the separation starts at order $\mu^2$ for small
$\mu$. The only case where this separation does not occur is for $L
= \pi R$ (i.e.~for $u_0 = u_\Lambda$), since then constant values of $x_4$
always solve the equations of motion.

The analysis in the intermediate temperature phase is similar. The
action is now
\begin{equation}
\label{interT} S = \frac{\hat{T}_8}{g_s} \int\!{\rm d}x_4\, u^4\,
     \sqrt{f + \left( \frac{R_{\text{D4}}}{u}\right)^{3}
     (u'{}^2 - F_{04}^2)} \, .
\end{equation}
The Hamiltonian takes the form
\begin{equation}
\label{HaintT} H = \frac{\displaystyle  u^4 f(u)}{\displaystyle
 \sqrt{f + \left( \frac{R_{\text{D4}}}{u}\right)^{3}
     (u'{}^2 - F_{04}^2)}} \, \, ,
\end{equation}
and the conserved $E$ takes the form
\begin{equation}
\label{eomA0intT} E \equiv \frac{\displaystyle u^4\,
F_{04}}{\displaystyle \left(\frac{u}{R_{\text{D4}}}\right)^3
 \sqrt{f(u) + \left( \frac{R_{\text{D4}}}{u}\right)^{3}
     (u'{}^2 - F_{04}^2)}} \, = \text{const}\,.
\end{equation}
Dividing~\eqref{eomA0intT} by~\eqref{HaintT} we get for the electric
field the same expression \eqref{e:EzeroT} as before, and the
expression for $H(u_0)$ is also identical to the previous expression
\eqref{forh}. Solving for $u'$ we now find
\begin{equation}
\label{upint} u' = \sqrt{ u^8 f(u) - u_0^8 f(u_0) +
  \frac{E^2}{R_{\text{D4}}^3} \big( f(u) u^3 - f(u_0) u_0^3 \big)}
  \left(\frac{u}{R_{\text{D4}}}\right)^{3/2}\,\sqrt{f(u)}\,
  \frac{1}{H}\,.
\end{equation}
The relations between the parameters now take the form :
\begin{equation}
\mu_A = - E \int_{u_0}^{\infty} {\rm d}u \sqrt{f(u)}
\left(\frac{u}{R_{D4}}\right)^{3/2} \left[ u^8 f(u) - u_0^8 f(u_0) +
  \frac{E^2}{R_{\text{D4}}^3} \big( f(u) u^3 - f(u_0) u_0^3 \big)
  \right]^{-1/2},
\end{equation}
\begin{equation}
L = 2 H \int_{u_0}^{\infty} \frac{{\rm d}u}{\sqrt{f(u)}}
\left(\frac{R_{D4}}{u}\right)^{3/2} \left[ u^8 f(u) - u_0^8 f(u_0) +
  \frac{E^2}{R_{\text{D4}}^3} \big( f(u) u^3 - f(u_0) u_0^3 \big)
  \right]^{-1/2}.
\end{equation}

\section{Instability of isotropic solutions and rho meson condensation}
\label{s:stability}
\subsection{The fluctuation equations}

In the previous section we found the static isotropic solutions of
the Sakai-Sugimoto model in the presence of flavor chemical potentials.
These solutions are the unique isotropic solutions in the absence of
sources, but there may exist additional non-isotropic solutions, which
could dominate the thermodynamics for some values of the chemical
potential. We will focus on the isospin chemical potential case. Since
the model contains massive particles charged under the isospin symmetry,
one might expect that for a large enough chemical potential these
particles will want to condense and modify the solution, and we will
see that this is indeed the case.

In order to analyze the stability in general we need to use the full
DBI action, and include also the Chern-Simons term in the action of
the five dimensional gauge field (which includes quadratic terms in
the fluctuations around the solutions of the previous
section). However, as discussed above, for $\mu \ll \lambda_5 / L^2$
the DBI action can be approximated just by the Yang-Mills action, and
in the same regime one can also show that the contribution of the
Chern-Simons term is much smaller than that of the Yang-Mills terms,
so it can also be ignored. Thus, in this section we will use the
Yang-Mills action to look for instabilities, and we will see that the
isotropic phase discussed in the previous section becomes unstable
already for $\mu$ which obeys $\mu \ll \lambda_5 / L^2$, so that the
analysis is self-consistent. For simplicity, we will consider in this
section only the low-temperature phase. We will also consider only the
special case of $L = \pi R$ for which the equations (and the isotropic
solution) simplify.  It should be straightforward to generalize our
analysis, but we postpone this to future work.

To analyze the stability we need to consider the spectrum of
fluctuations of all the fields in the theory around our solutions. In the
limit of weak curvature, we can limit our analysis to the massless fields
living on the D8-branes, since the other mesons are much heavier.
Because the chemical potential breaks Lorentz symmetry, there are four
types of fields which have to be discussed: there are KK towers of
scalars, transverse vectors, and longitudinal vectors, and in addition
there are the pions.  All of these except the scalars arise from the
five-dimensional vector field. In order to keep the vector equations
of motion reasonably compact, we will temporarily revert back to
the~$u$ coordinate system. The action (in the approximation described
above) then reads
\begin{equation}
\label{e:SplusCS}
S =  \tilde{T} \frac{3 R_{D4}^3}{8 u_0} \! \int\!{\rm d}^4x\,{\rm d}u\,\Big[
u^{-1/2} \gamma^{1/2} \Tr(F_{\mu\nu} F_{\rho\sigma}) \eta^{\mu\rho}\eta^{\nu\sigma}
+ 2 R_{D4}^{-3} u^{5/2} \gamma^{-1/2} \Tr(F_{\mu u} F_{\nu u})\eta^{\mu\nu}\big]\,.
\end{equation}
As in the previous section, we will analyze the effect of a vectorial
isospin density in the~$U=U_{\text{min}}$ vacuum where the pions have
condensed by performing a global symmetry transformation taking this
back to the~$U=I$ vacuum, in which the isospin chemical potential
becomes an axial isospin chemical potential.  So, we will analyze the
stability of the solution constructed in \S\ref{s:gfsols} (which we write
here just for the D8-brane branch $z>0$),
\begin{equation}
\label{e:backgroundA0}
\bar{A}_0 =
\mu_I \frac{\sigma_3}{2}\, \bar{A}(u)\,,\quad
\bar{A}(u)\equiv {\cal N}\int_{u_0}^{u}\!\frac{{\rm d}\tilde{u}}{{\tilde u}^{5/2}
  \,\gamma(\tilde{u})^{-1/2}}
\qquad\Rightarrow\qquad
\bar{F}_{0u} = -\mu_I \frac{\sigma_3}{2} \frac{{\cal N}}{u^{5/2} \gamma^{-1/2}}\,,
\end{equation}
with ${\cal N}$ chosen to make~$\lim_{u\rightarrow\infty} \bar{A}(u) =
1$.  We will continue to use the gauge $A_u=0$. In this gauge the
expansion of the gauge field in terms of fluctuations is
\begin{equation}
A_0 = \bar{A}_0(u) + \delta A_0^{(a)}(\omega,\vec{k},u)\,\sigma_a\,e^{i\omega t+i\vec{k}\cdot\vec{x}}\,,\quad
A_i = \delta A_i^{(a)}(\omega,\vec{k},u)\,\sigma_a\,e^{i\omega
  t+i\vec{k}\cdot\vec{x}}\,,\quad
A_u = 0\,.
\end{equation}
The transverse fields coming from~$\delta A_i(\omega,\vec{k},u)$ will
be degenerate since there is still rotational invariance, but they
need no longer be degenerate with $\delta A_0$.

The equations of motion for the diagonal gauge field fluctuations~$\delta
A_i^{(0,3)}$ and $\delta A_0^{(0,3)}$ do not change upon inclusion of
the chemical potential and the background \eqref{e:backgroundA0}; 
their spectra remain unmodified (in the approximation of $\mu_I \ll \lambda_5 / L^2$
where we ignore the Chern-Simons term in the action). For the other
matrix elements, the action~\eqref{e:SplusCS} leads to the following
equations for the spatial components,
\begin{align}
\label{eq3}
& R_{D4}^{-3} u^{1/2} \gamma^{-1/2} \partial_u\Big( u^{5/2} \gamma^{-1/2}
\partial_u  \delta A_i^{(1)} \Big) 
+   \partial^2 \delta A_i^{(1)}  -
\partial_i (\partial_\mu \delta A^{\mu (1)})\nonumber\\
& \hspace{8em} + 2 \mu_I \bar{A}(u) \partial_0 \delta A_i^{(2)}
-  \mu_I \bar{A}(u) \partial_i \delta A_0^{(2)} =
- \mu_I^2 \bar{A}(u)^2 \delta A_i^{(1)}\,,  \\[1ex]
\label{eq4}
& R_{D4}^{-3} u^{1/2} \gamma^{-1/2} \partial_u\Big( u^{5/2} \gamma^{-1/2}
\partial_u  \delta A_i^{(2)} \Big)
 +  \partial^2 \delta A_i^{(2)}  -
\partial_i (\partial_\mu \delta A^{\mu (2)}) \nonumber \\
&\hspace{8em} - 2 \mu_I \bar{A}(u) \partial_0 \delta A_i^{(1)}
+ \mu_I \bar{A}(u) \partial_i \delta A_0^{(1)} =
- \mu_I^2 \bar{A}(u)^2 \delta A_i^{(2)} \,.
\end{align}
Here~$\partial^2 \equiv \partial_\mu \partial^\mu \rightarrow \omega^2
- \vec{k}^2$ and~$F_{\mu\nu} = \partial_\mu A_\nu - \partial_\nu A_\mu
- i [ A_\mu, A_\nu]$, so that with our convention~$A_\mu =
A_\mu^{(a)}\sigma_a$ we have $F^{(a)}_{\mu\nu} = \partial_\mu A^{(a)}_\nu - \partial_\nu A^{(a)}_\mu
+ 2 \epsilon^{abc} A_\mu^{(b)} A_\nu^{(c)}$.
In addition, we have the equations for the
time-like and $u$-components of the gauge field, of the form
\begin{align}
\label{eq0}
& R_{D4}^{-3} u^{1/2} \gamma^{-1/2} \partial_u(u^{5/2} \gamma^{-1/2} \partial_u
\delta A^{(1)}_0) \nonumber\\[1ex]
&\hspace{.2\textwidth}
+  \bigg{(}\partial_i \partial^i \delta A^{(1)}_0 - \partial_0
(\partial_i \delta A_i^{(1)}) + \mu_I \partial_i \delta A_i^{(2)} \bar{A}(u) \bigg{)}  = 0\,,\\
& R_{D4}^{-3} u^{1/2} \gamma^{-1/2} \partial_u(u^{5/2} \gamma^{-1/2} \partial_u
\delta A^{(2)}_0) \nonumber\\[1ex]
&\hspace{.2\textwidth} +
 \bigg{(}\partial_i \partial^i \delta A^{(2)}_0 - \partial_0
(\partial_i \delta A_i^{(2)}) - \mu_I \partial_i \delta A_i^{(1)} \bar{A}(u) \bigg{)}  = 0\,,\\
\label{equ}
&\partial_u (\partial_0 \delta A^{(1)}_0 -
\partial_i \delta A^{(1)}_i)  - \mu_I \bar{A}(u) \partial_u \delta
A^{(2)}_0 + \mu_I \delta A_0^{(2)}\bar{A}'(u) = 0 \, ,\\
\label{equ2}
&\partial_u (\partial_0 \delta A^{(2)}_0 -
\partial_i \delta A^{(2)}_i)  + \mu_I \bar{A}(u) \partial_u \delta
A^{(1)}_0 - \mu_I \delta A_0^{(1)}\bar{A}'(u) = 0 \, .
\end{align}
The equations of motion for the scalar fields will be discussed in the
next subsection.

Naively, one may expect that for positive $\mu_I$ the masses of the fields
carrying positive isospin charge would go down and the masses of the fields
carrying negative isospin charge would go up (at least for small $\mu_I$).
The equations above are indeed symmetric under $\mu_I \to -\mu_I$, $A^{(2)} \to
-A^{(2)}$ which changes the sign of the isospin chemical potential and also
exchanges the positively and negatively charged mesons. However, there is
an extra symmetry of the equations (and of the solution \eqref{e:backgroundA0})
under \mbox{$\mu_I \to -\mu_I$} together with \mbox{$z\to
-z$} (when the equations are written in the $z$ coordinate), which implies
that in fact the spectrum is the same for positive and for negative $\mu_I$
(note that together with the previous symmetry, this implies that the spectrum is
the same for the positively charged mesons as for the negatively charged
mesons). The reason for this is the condensation of the pions; the
expectation above holds when we expand around the trivial vacuum $U=I$, but
it does not hold when we expand around the vacuum in which the pions have
condensed (or, equivalently, when we expand around the vacuum $U=I$ with
an axial isospin chemical potential). Note in particular that the
mesons that we analyze in this section carry a $U(1)$ charge in the new
vacuum, but this is not the same as the original $U(1)$ isospin charge to
which the chemical potential was coupled\footnote{In the original variables,
if we choose $U_{\text{min}} \propto \sigma_1$, the equations above involve the
$\sigma_1$ and $\sigma_3$ components of the matrices.}. In any case, we will see that the
masses of some of the fields (of both positive and negative $U(1)$ charge)
go down with $\mu_I$, and that eventually they condense. Because
of the symmetries discussed above, it is enough to consider positive values of $\mu_I$,
and to consider either the positively charged or the negatively charged
mesons (but not both), and we will use this when solving the equations.

\subsection{Transverse vectors and scalars}

\begin{figure}[t]
\begin{center}
\includegraphics[width=.95\textwidth]{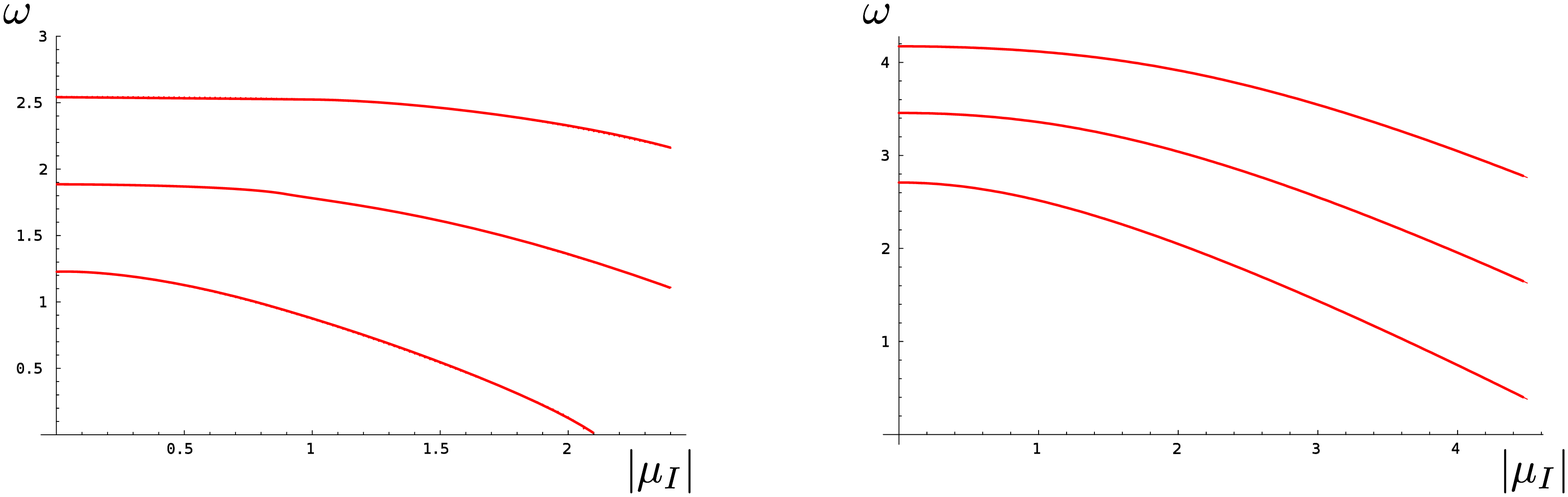}
\end{center}
\caption{The frequencies of zero spatial momentum solutions
  for the transverse vectors (left)
  and scalars (right), as a function of the isospin chemical
  potential, in units of~$\sqrt{u_{\Lambda}/R_{\text{D4}}^3}=2/(3R)$.
The frequencies at $\mu_I=0$ reproduce the spectrum found in \cite{Sakai:2004cn}.
\label{f:transverse_and_scalars}}
\end{figure}

The analysis of the transverse vectors and the scalars is similar to
their analysis without the chemical potential~\cite{Sakai:2004cn}. Let
us first focus on the solutions of the~$\delta A_i^{(1,2)}$
fluctuations, which describe massive transverse vector mesons. For
these modes, it is consistent to set~$\delta A_0=0$ and then also
impose the standard constraint for massive transverse
vectors~$\partial_i \delta A^i=0$. This automatically takes care of
equations~\eqref{eq0}-\eqref{equ2}.  The spectrum of
eigenfunctions~$A_i^{(1,2)}$ of~\eqref{eq3} and~\eqref{eq4} and the
corresponding eigenvalues~$\omega$ can be found numerically, using
standard shooting methods, as a function of~$\mu_I$. For the special
case of $L = \pi R$ they are displayed on the left-hand side of
figure~\ref{f:transverse_and_scalars}, in units of
$\sqrt{u_{\Lambda}/R_{D4}^3} = 2/(3R)$.  We see that the frequency of the lowest
transverse vector goes to zero at
a chemical potential of magnitude~$\mu_{\text{crit}} \approx
2.1*2/(3R)\approx 1.7\, m_{\rho}$, indicating a potential instability there.

A similar story to the one above holds for the scalar mesons, arising
from the scalar field $y$ on the D8-brane\footnote{There are additional
  scalar fields coming from the components of the gauge field along
  the $S^4$, but they are heavier in the absence of the chemical
  potential, so we will assume that they do not condense and ignore
  them in our analysis.}.  The scalar mesons feel the chemical
potential through the commutators in the covariant derivatives. The
equation of motion for the scalars in the $I$ and $\sigma_3$
directions are unaffected by the chemical
potential~\eqref{e:backgroundA0}, while the equations for
the~$\sigma_1$ and $\sigma_2$ components are easily generalized from
those in~\cite{Sakai:2004cn}. In the~$z$-coordinate system they read
(for $L = \pi R$)
\begin{equation}
\frac{1}{R^2} K^{1/3}\Big[ u_{\Lambda}^2 \partial_z\Big( K \partial_z y^{(1,2)} \Big) - 2
  y^{(1,2)}\Big]  + (\omega^2 - k^2)\, y^{(1,2)} =
\pm 2i\mu_I \omega \bar{A} y^{(2,1)}
- \mu_I^2\,\bar{A}^2 y^{(1,2)}\, ,
\end{equation}
where $K(z) \equiv (u/u_{\Lambda})^3$, and
where the plus/minus sign refers to the $y^{(1)}$ and $y^{(2)}$
functions, respectively. These equations are diagonal in the
basis of positively and negatively charged mesons, given by
\begin{equation}
y^{(1)} = \pm i y^{(2)}\,.
\end{equation}
The resulting behavior of the~$\omega$ eigenvalues as a function of
chemical potential is displayed on the right-hand side of
figure~\ref{f:transverse_and_scalars}. We observe that the scalars
also go to zero frequency leading to potential instabilities, 
but only for chemical potentials which are
larger in magnitude than those for which the first vector meson
becomes unstable, so the vector instability must be analyzed first.

\subsection{Pions and longitudinal vectors}
\label{pionvects}

Let us now turn our attention to the pions and longitudinal
vectors. For any non-zero momentum, these particles carry the same
quantum numbers under the rotation symmetry preserved by the chemical
potential, so it is possible for them to mix as the chemical potential
is increased. Both the pions and the longitudinal vector excitations
involve the two components
\begin{equation}
A^{(1,2)}_i = i k_i A^{(1,2)}_T\,,\quad\text{and}\quad A^{(1,2)}_0\,.
\end{equation}
Again, the equations are diagonal when we choose
\begin{equation}
\label{e:pionansatz}
\delta A_i^{(1)} = \pm i \delta A_i^{(2)} \,,\qquad
\delta A_0^{(1)} = \pm i \delta A_0^{(2)}\,.
\end{equation}
We will use the polarization corresponding to the plus signs below,
and keep~$\delta A_0^{(2)}$ and $\delta A_i^{(1)}$ as the two independent
components (with the others given by \eqref{e:pionansatz}); as discussed
above, the other polarization gives identical results. 

The boundary conditions determine how the pions and longitudinal
vectors are encoded in these fields. The general solution to the
equations of motion of the two functions~$A_T$ and
$A_0$ near the boundary takes the form
\begin{equation}
\label{e:genericasymptote}
\delta A^{(1)}_T  = c_0 + \frac{c_1}{z} + \cdots\,,\qquad
\delta A^{(2)}_0  = d_0 + \frac{d_1}{z} + \cdots\,.
\end{equation}
Here~$c_0$ and $d_0$ are the coefficients of the non-normalizable modes,
while~$c_1$ and $d_1$ are the coefficients of the normalizable modes. The equation
of motion~\eqref{equ} imposes one constraint on these four free
parameters, thereby leaving a two-parameter family after the
overall normalization is fixed.

The solutions for the pions are expected to have a non-zero value of
$\delta A_T$ at $z \to \pm \infty$ (as for vanishing chemical
potential in the $A_u=0$ gauge), and we normalize this value to
$\pi/2$. On the other hand, the gauge field $F_{0i}$ must vanish as $z
\to \pm \infty$, for the solution to have finite action.
This constraint determines~$d_0$ in terms of $c_0$, leaving
\begin{equation}
\label{e:pionasymptote}
\text{pion}:\qquad\left\{
\begin{aligned}
\delta A^{(1)}_T(z\rightarrow\infty) &= \frac{\pi}{2} + \frac{c_1}{z}+
\ldots\,,\\[1ex]
\delta A^{(2)}_0(z\rightarrow\infty) &= (\omega + \mu_I) \frac{\pi}{2} +
\left(\frac{c_1\, k^2}{\omega + \mu_I} - \mu_I\right) \frac{1}{z}+\ldots\,.
\end{aligned}\right.
\end{equation}
Note that as $\mu_I \to 0$ we have
$\delta A_0^{(2)} = \omega \delta A_T^{(1)}$, as expected due to the enhanced
Lorentz symmetry there. 

On the other hand, the solutions corresponding to longitudinal vector
mesons are characterized by the absence of
non-normalizable components in~$A_i$ and $A_0$, so that we have (in an arbitrary
normalization)
\begin{equation}
\label{e:longasymptote}
\text{longitudinal vectors}:\qquad\left\{
\begin{aligned}
\delta A^{(1)}_T(z\rightarrow\infty) &= \frac{1}{z} + \ldots\,,\\[1ex]
\delta A^{(2)}_0(z\rightarrow\infty) &= \frac{k^2}{\omega + \mu_I} \frac{1}{z}+\ldots\,.
\end{aligned}\right.
\end{equation}
Note that if~$c_1\rightarrow\infty$ in \eqref{e:pionasymptote}, the pion modes look like
longitudinal vector modes (with infinite overall normalization); this
will be important for us later.

\begin{figure}[t]
\begin{center}
\includegraphics[width=.6\textwidth]{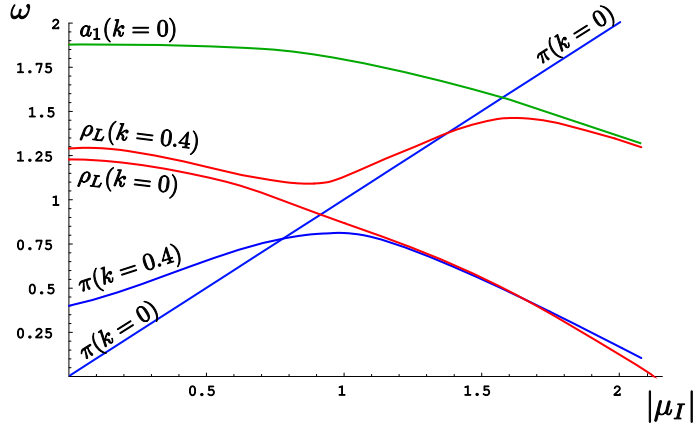}
\end{center}
\caption{The energy~$\omega$ as a function of the isospin chemical
  potential~$\mu_I$, in units of $\sqrt{u_{\Lambda}/R_{D4}^3}=2/(3R)$,
  for the sector consisting (at $\mu_I=0$) of the pion~$\pi$ and
  the first two longitudinal vector mesons~$\rho_L$ and $a_1$. At~$\mu_I\approx
  2.1*2/(3R)$ the first longitudinal vector meson becomes unstable. This
  coincides with the instability of the transverse vector meson
  depicted in figure~\protect\ref{f:transverse_and_scalars}.\label{f:pion_w_vs_mu}}
\end{figure}

Let us now discuss the relation between the boundary conditions
at~$z=\pm\infty$. At~$\mu_I=0$ the pion is encoded in~$A_T^{(1)}$ which
is anti-symmetric in $z$; the relation of the pion to a chiral
symmetry transformation implies that this should remain true near the
boundary also for non-zero~$\mu_I$,
\begin{equation}
A_T^{(1)}(z=+\infty) = - A_T^{(1)}(z=-\infty)\,.
\end{equation}
For the zeroth component, the boundary condition comes from
requiring $F_{0i}=0$ at the boundary, which gives
\begin{equation}
\label{e:bdyA0i}
\begin{aligned}
A_0^{(2)}(z=+\infty) &= (\omega + \mu_I)\, A_T^{(1)}(z=+\infty)\,,\\[1ex]
A_0^{(2)}(z=-\infty) &= (\omega - \mu_I)\, A_T^{(1)}(z=-\infty) =
(-\omega + \mu_I) A^{(1)}_T(z=+\infty)\,.
\end{aligned}
\end{equation}

Before we discuss the states at generic values of the momentum~$k$, let
us first analyze the~\mbox{$k=0$} case. This case can in fact be
solved analytically, again specializing to the case of $L = \pi R$,
where ${\bar A}(z) = 2 \arctan(z/u_{\Lambda}) / \pi$. The equation of motion~\eqref{equ} becomes
\begin{equation}
-2\mu_I \delta A_0^{(2)} + u_{\Lambda}(1+z^2/u_{\Lambda}^2)
\Big( (\pi\omega+2\mu_I \arctan (z/u_{\Lambda})) \partial_z
\delta A_0^{(2)}(z)\Big) = 0\,.
\end{equation}
This is solved by
\begin{equation}
\delta A_0^{(2)} = (\omega + \frac{2\mu_I}{\pi}\arctan (z/u_{\Lambda})) \cdot \text{const.}
\end{equation}
In order to make this satisfy the boundary conditions~\eqref{e:bdyA0i}
we are forced to choose~$\omega = \mu_I$. This simple~$k=0$ state
becomes the pion in the limit~$\omega,\mu_I\rightarrow 0$. We see that its
frequency goes up with $|\mu_I|$ and never vanishes.

In order to analyze the fate of the remainder of the spectrum in the
pion/longitudinal vector sector, we resort to a numerical solution of
the equations of motion. Modes satisfying~\eqref{e:longasymptote} are
found in the usual way with a one-parameter shooting method, starting
from $z=\infty$ and looking for values of $\omega$ such that the solution
goes to zero as $z \to -\infty$. For the
modes which are pions at~$\mu_I=0$, i.e.~modes
satisfying~\eqref{e:pionasymptote}, a search through the
two-dimensional space spanned by~$c_1$ and $\omega$ is required. Some
details about this procedure can be found in the appendix. 

The upshot of this analysis is the following. The
modes which start out as longitudinal vectors when $\mu_I=0$ all have
decreasing~$\omega$ as $\mu_I$ is increased, and the frequency of the
lowest one goes to zero at $\mu_I=\mu_{\text{crit}}$.
For the modes starting out with pion boundary conditions, the
situation is more subtle. They start by going up, but they mix with
the other modes. The low-momentum modes starting out as pions develop a
large value for~$c_1$, which goes to infinity as~$\mu_I\rightarrow
\mu_{\text{crit}}$. These modes therefore go over from
pure pions to pure longitudinal vectors\footnote{Recall that to study the
pions we arbitrarily fixed $c_0=\pi/2$, but a more appropriate normalization
when $c_1 \to \infty$ is $c_0 \to 0$ with $c_1$ remaining finite.}
as $\mu_I$ goes from zero to
$\mu_{\text{crit}}$. Similarly, the longitudinal vectors develop a
pion component, and as $\mu_I \to \mu_{\text{crit}}$ one linear
combination of the modes which starts out as longitudinal vectors
may be identified with the pion.

In figure~\ref{f:pion_w_vs_mu} we plot the value of~$\omega$ for the
pion and longitudinal vector states for a range of values of~$\mu_I$, at
fixed~$k$. This plot first of all shows that a small chemical
potential leads to massive pions with $\omega^2 = k^2 + \mu_I^2$,
in accordance with our analysis of the 
chiral Lagrangian in the previous section. 
As $\mu_I$ is increased, the frequencies of the zero-momentum pions and
of the lightest longitudinal vectors cross each other, and the lightest state becomes a
longitudinal vector rather than a pion. Note that for non-zero $k$ the two states cannot
cross each other since they then have the same quantum numbers, but instead
they mix as described above, and they come closer and closer to level-crossing
as $k \to 0$.

For a value of~$\mu_I \approx 2.1*2/(3R)$, the lowest-mass
longitudinal vector meson (which we call the ``rho'') goes to zero
frequency (at zero spatial momentum), indicating a potential instability. This is
precisely the point where we also found that the lowest-mass transverse
vector meson becomes unstable. Thus, the conclusion is that an
instability of the rho meson develops at~$\mu_{\text{crit}} \approx 2.1*2/(3R) 
\approx 1.7\,m_{\rho}$, where~$m_{\rho}$ is the mass of the rho meson at zero
chemical potential. The fact that the transverse vectors and longitudinal vectors go to
zero frequency at the same value of $\mu_I$ is not surprising, since for zero
spatial momentum (where we find the first instability) there is no
distinction between transverse and longitudinal vectors.

Note that the Skyrme model, which (as mentioned above) is a good
approximation to the low-energy physics of the Sakai-Sugimoto model,
indicates that the first instability (the first mode to go to
zero frequency) as we increase $\mu_I$ comes from
pions with $\mu_I$ and $k$ of the order of the scale of the Skyrme term
(which is of the same order as the rho meson mass in the Sakai-Sugimoto
model). This differs from our conclusion that
the first instability comes from vector mesons of zero momentum. Of course,
this is not a contradiction, since the Skyrme approximation breaks
down when $\mu_I$ and/or $k$ become of order the mass of the vector mesons, and the
additional vector fields must be included in the analysis then (as
we did in our analysis).

\subsection{Vector meson condensation}

We have seen that at~$\mu_I\approx 1.7\, m_{\rho}$, the zero momentum
modes of the ``rho'' vector meson go to zero frequency. This suggests
that the theory becomes unstable there towards forming a condensate of
rho mesons, in addition to the pion condensate which is already
present; this was conjectured to happen also in
QCD~\cite{Voskresensky:1997ub,Lenaghan:2001sd,Sannino:2002wp}.  This
condensate is of a vector field, so it breaks the rotational symmetry
$\text{SO}(3)\rightarrow \text{SO}(2)$, and since it involves the
$\sigma_1$ and/or $\sigma_2$ components of the meson, it also
completely breaks the remaining $U(1)$ flavor symmetry.\footnote{There
  is another possible condensate of the vector meson, in which the
  unbroken~U(1) is a diagonal subgroup of the~U(1) flavor symmetry and
  an~SO(2) rotational symmetry~\cite{Sannino:2002wp}. Without loss of
  generality, we can assume that in this phase the non-zero components
  of the gauge field are \mbox{$A_1^{(1)} = A_2^{(2)}$} (in addition to
  $A_0^{(3)}$). We have verified numerically that in our model this
  alternative condensate always has higher energy than the condensate
  discussed below, so we will not consider it further.}

In order to find this new ground state, we assume without loss of
generality that the third component of the vector rho meson condenses,
and that the condensate is proportional to $\sigma_1$. We also assume
a static and translationally-invariant solution.
In this convention, the Goldstone
bosons for the broken rotational symmetry are the fluctuations
of~$A_2^{(1)}$ and~$A_1^{(1)}$, while the fluctuations of~$A_3^{(2)}$ include the
Goldstone boson for the broken flavor~$U(1)$.
We then have to solve the non-linear equations of motion
\begin{equation}
\label{e:nonlinear}
\begin{aligned}
R_{D4}^{-3} \partial_u \left[ u^{5/2}\gamma^{-1/2} \partial_u A_0^{(3)} \right] &= \phantom{-}
4 (A_3^{(1)})^2 A_0^{(3)}\,u^{-1/2}\gamma^{1/2}\,,\\[1ex]
R_{D4}^{-3} \partial_u \left[ u^{5/2}\gamma^{-1/2} \partial_u A_3^{(1)} \right] &= -
4 (A_0^{(3)})^2 A_3^{(1)}\,u^{-1/2}\gamma^{1/2}\,,
\end{aligned}
\end{equation}
with the boundary conditions $A_0^{(3)}(z=\pm \infty) = \pm \mu_I/2$, $A_3^{(1)}
(z=\pm \infty)=0$.
We expect that for~$\mu_I < \mu_{\text{crit}}$ these equations admit
only the solution of the previous section which describes the pion condensate, i.e.
\begin{equation}
\mu_I < \mu_{\text{crit}}:\qquad A_0^{(3)} = \frac{\mu_I}{\pi}
\arctan (\frac{z}{u_{\Lambda}})\,,\quad A_3^{(1)} = 0\,.
\end{equation}
For~$\mu_I > \mu_{\text{crit}}$ we expect the appearance of an additional solution, which has both
~$A_0^{(3)}$ and $A_3^{(1)}$ non-zero.
\begin{figure}[t]
\begin{center}
\includegraphics[width=.4\textwidth]{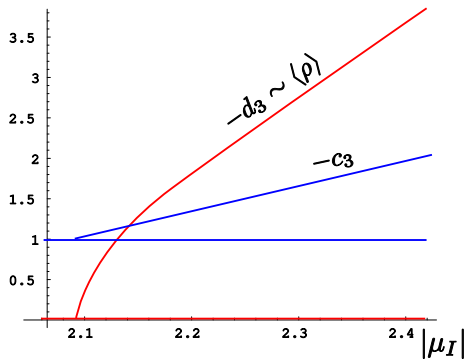}
\end{center}
\caption{Appearance of the new ground state in which the $\rho$ meson
  has condensed. This new ground state also still has a non-zero pion
  condensate. We plot $(-c_3)$ (in units of $u_{\Lambda}$) and
  $(-d_3)$ (in units of $(u_{\Lambda}/R_{D4})^{3/2}$) for both solutions, 
  where $c_3$ and $d_3$ are 
  defined by
$A_0^{(3)} = \mu_I (1/2 + c_3 / \pi z + \cdots)$, $A_3^{(1)} = d_3/z + \cdots.$}
\label{condensate_fig}
\end{figure}

We again resort to a numerical analysis in order to find the
solution. Details of this procedure can be found in the appendix. The
upshot is that indeed a rho meson condensate forms for~$\mu_I>
\mu_{\text{crit}}$, so there is a second order phase transition there. 
Around $\mu_I=\mu_{\text{crit}}$ this condensate is, to good approximation, given
by
\begin{equation}
\langle\rho\rangle \propto \begin{cases}
0                            & \text{for $\mu_I < \mu_{\text{crit}}$,}\\[1ex]
\displaystyle\sqrt{\mu_I-\mu_{\text{crit}}} & \text{for $\mu_I \geq \mu_{\text{crit}}$.}
\end{cases}
\end{equation}
This condensate is drawn in figure \ref{condensate_fig}, using an arbitrary
normalization in which we read it off from the coefficient of the normalizable
mode of $A_3^{(1)}$; it should be possible to translate this to
a physical normalization where the meson has a canonical kinetic term, but we have
not done this here, and we do not expect this to make a big difference near the
phase transition. The value of $(-\mu_I*c_3)$ in this figure is the isospin
charge density (up to a constant); note that in the solution with the rho condensate it rises
more rapidly than in the solution that has only the pion condensate.

The fact that the expectation value of the rho meson scales as 
$\sqrt{\mu_I-\mu_{\text{crit}}}$ near the transition is expected
on general grounds, by a Landau-Ginzburg type analysis of the rho meson
effective theory, assuming that the effective mass squared of
the rho mesons near the transition is proportional to $(\mu_{\text{crit}}
-\mu_I)$, and that there is an effective potential of fourth order in the
rho mesons which stabilizes the condensate. Our analysis implies
that the coefficient of this fourth order term is positive.

\subsection{Potential generalizations}

As indicated above, our stability analysis so far is limited to the special case
of $L = \pi R$, and it would be interesting to generalize it to other
values of $L/R$, and to see whether the instability of the isotropic
phase persists for all values of $L/R$, and how $\mu_{\text{crit}}$
depends on this parameter. For small enough values of $L/R$ there is
also an intermediate temperature phase in which the branes are
connected~\cite{Aharony:2006da}, and it would be interesting to see if
a similar instability arises in that phase as well or not. In that
phase there are also deconfined quarks in the spectrum, which may play
an interesting role.

Our analysis so far is limited to the first instability which arises
as we increase $\mu_I$, which leads to a condensation of the lowest
vector meson. It would be interesting to see what happens as $\mu_I$ is
increased further -- do additional fields also want to condense?
This analysis is more difficult than the analysis we performed here
since the solutions for $\mu_I > \mu_{\text{crit}}$ which one has to
expand around are only known numerically, but still it should be
possible to do it. Of course, eventually, as we increase $\mu_I$ to
order $\lambda_5 / L^2$, we also need to include the DBI and
Chern-Simons terms in the action. Baryons may also eventually become
relevant for the analysis (recall that in the Sakai-Sugimoto model,
the ratio of the baryon mass to its isospin scales as $\lambda_5/R$
compared to that of the mesons~\cite{Sakai:2004cn,Hata:2007mb}, so
baryons are not expected to be relevant for small~$\mu_I$). It would
also be interesting to look for possible non-translationally-invariant
solutions.

Naively, our analysis in this section implies that the large $\mu_I$
solutions of \S\ref{s:gfsolslarge} are never relevant,
since the isotropic solutions found in that section 
do not dominate already for smaller
values of $\mu_I$. However, there are several situations where these
solutions may still be relevant. One is the case of an axial $U(1)$
chemical potential (which is also part of the full phase diagram of
QCD, and may be interesting in particular at large $N_c$ where the
axial symmetry is approximately conserved). In this case there are 
no instabilities for small values of $\mu$ (for which the spectrum
is unmodified), but the Chern-Simons
term may well lead to instabilities at larger values of $\mu$ (of
order $\lambda_5/L^2$),
similar to those discussed
in~\cite{Domokos:2007kt}. To see this one has to expand the action
of a single D8-brane around the
solutions of \S\ref{s:gfsolslarge}, using the full DBI plus Chern-Simons
action. Another case where these solutions could be relevant is
the case of $N_f > 2$ with generic $SU(N_f)$ isospin chemical
potentials, with all eigenvalues different. When $L < \pi R$ this
leads to a separation of the D8-branes, which gives a mass to the
off-diagonal matrix elements, and this may eliminate the
instabilities discussed in this section which come from off-diagonal
elements of the meson field. Again, new instabilities
may then arise for $\mu \sim \lambda_5 / L^2$; it would be interesting
to investigate this further.

\vfill\vfill\vfill
\section*{Acknowledgements}

We thank Philippe de Forcrand, Simon Hands, Andreas Karch, David Mateos,
Owe Philip\-sen, Francesco
Sannino, Tho\-mas Sch\"afer, Dam T. Son, Kim Splittorff, Misha
Stephanov, Arkady Vainshtein and Larry Yaffe for 
discussions, and the organizers of the Newton Institute programme
``Strong Fields, Integrability and Strings'' and the workshop on
``Exploring QCD: Deconfinement, Extreme Environments and Holography''
for hospitality and for
creating a stimulating environment. OA would also like to thank Stanford
University, SLAC and Tel-Aviv University for hospitality during the course 
of this project. 

The work of OA and JS is supported in part by a center of excellence
supported by the Israel Science Foundation (grant number 1468/06), by
a grant (DIP H52) of the German Israel Project Cooperation, and by the
European Network MRTN-CT-2004-512194. The work of OA is also supported
in part by the Israel-U.S. Binational Science Foundation and by a
grant from the G.I.F., the German-Israeli Foundation for Scientific
Research and Development. The work of OA at the Newton Institute was
supported by a Senior Visiting Fellowship funded by EPSRC grant
531174. The work of KP was supported by VIDI grant 016.069.313 from
the Dutch Organization for Scientific Research (NWO).

\eject
\section{Appendix: physical modes}

In this appendix we provide some more technical details related to the
determination of the meson spectrum, as well as
the construction of the new ground state in which the rho meson is
condensed. This analysis is, in both cases, essentially based on a
standard shooting method for the two-point boundary value
problem. However, there are a few interesting features which we would
like to highlight.

\begin{figure}[b]
\begin{center}
\vspace{2ex}
\hbox{\vbox{\hbox{\includegraphics[width=.32\textwidth]{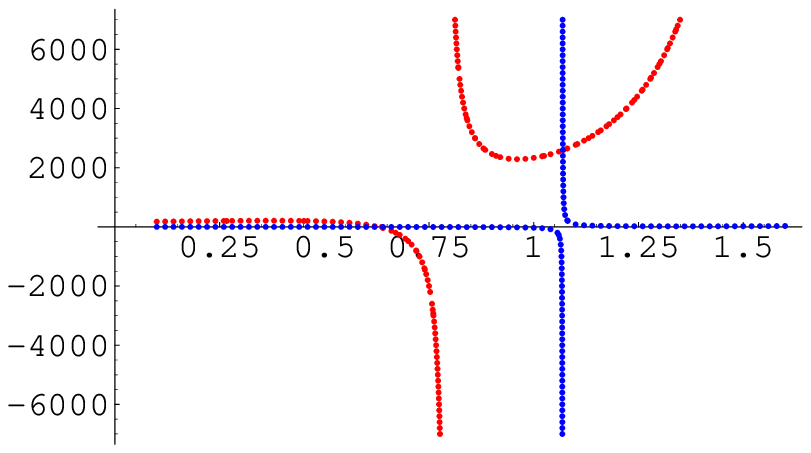}}\hbox{$\hspace{5em}\mu_I=0.63$}}
\vbox{\hbox{\includegraphics[width=.32\textwidth]{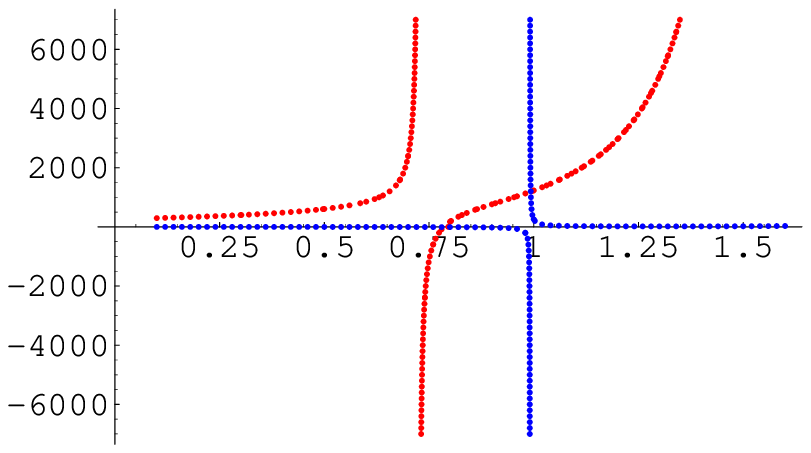}}\hbox{$\hspace{5em}\mu_I=0.79$}}
\vbox{\hbox{\includegraphics[width=.32\textwidth]{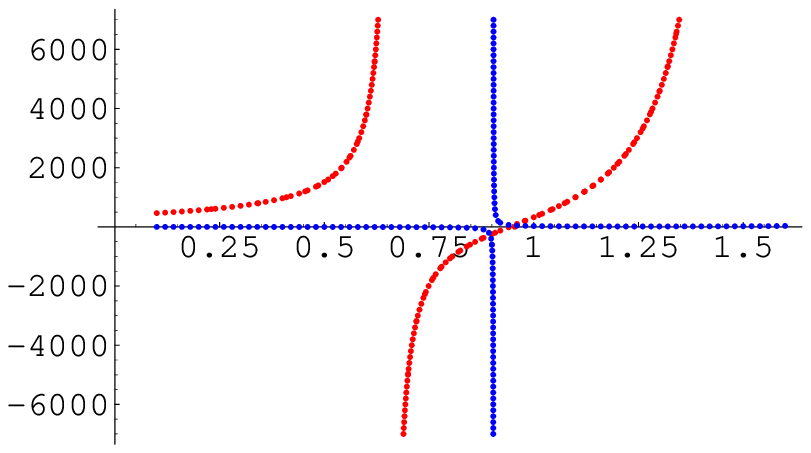}}\hbox{$\hspace{5em}\mu_I=0.94$}}}
\vspace{2ex}
\hbox{\vbox{\hbox{\includegraphics[width=.32\textwidth]{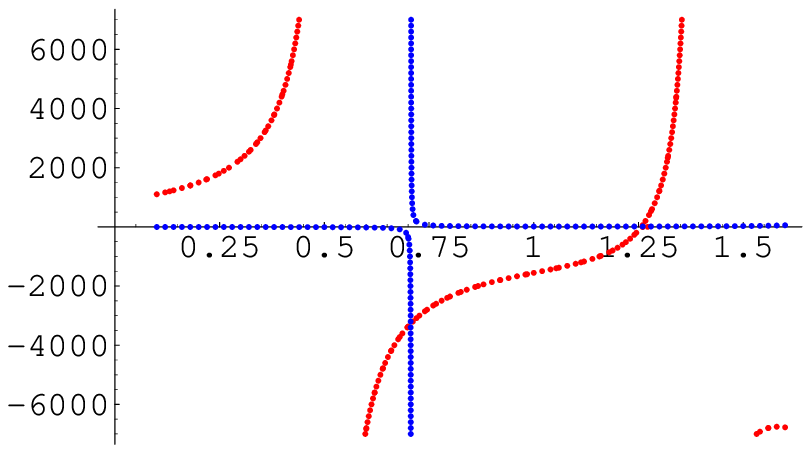}}\hbox{$\hspace{5em}\mu_I=1.26$}}
\vbox{\hbox{\includegraphics[width=.32\textwidth]{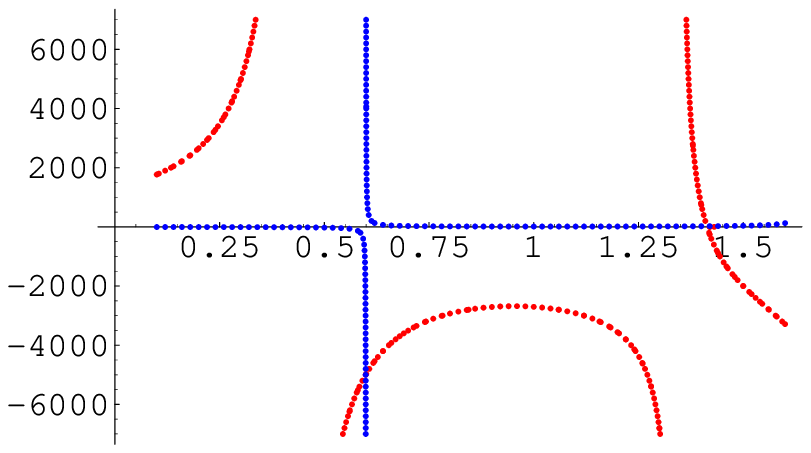}}\hbox{$\hspace{5em}\mu_I=1.41$}}
\vbox{\hbox{\includegraphics[width=.32\textwidth]{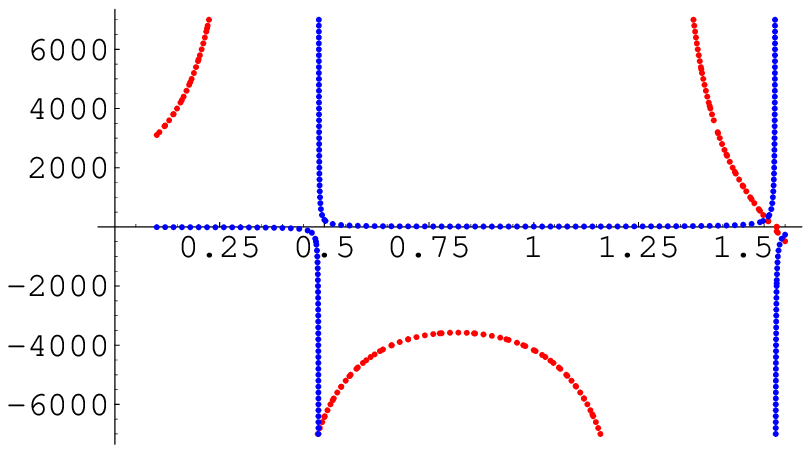}}\hbox{$\hspace{5em}\mu_I=1.57$}}}
\vspace{-4ex}
\end{center}
\caption{The location of the physical modes for small and fixed~$k$ ($k=0.05 \sqrt{u_{\Lambda}/R_{D4}^3}$)
  for various values of~$\mu_I$ (in units of $\sqrt{u_{\Lambda}/R_{D4}^3}=2/(3R)$).
  The horizontal and vertical axes label~$\omega$ (in the same units) and
  the coefficient of the $1/z$ term in $A_T^{(1)}$ (in units of $u_{\Lambda}$, in the
  normalization of \protect\eqref{e:pionasymptote}), respectively. The blue, almost
  straight curves indicate where the boundary condition
  on~$A_T^{(1)}(z=-\infty)$ is satisfied, while the red curves do the
  same for~\mbox{$A_0^{(2)}$}.
}
\label{f:rootplots}
\end{figure}

Let us first discuss the construction of solutions with asymptotic
expansion~\eqref{e:pionasymptote} and boundary conditions~\eqref{e:bdyA0i}.
When solving for the values of~$c_1$ and $\omega$ for which such a
mode exists, by shooting from $z=\infty$ to $z=-\infty$,
it is instructive to plot two curves, one for which
the~$A_0^{(2)}$ boundary condition is satisfied at $z=-\infty$
and one for which
the~$A_T^{(1)}$ boundary condition at $z=-\infty$ holds true. This leads to
figure~\ref{f:rootplots}. The intersection points of these two types
of curves indicate physical states. The intersection which is almost
on the $\omega$-axis has a $1/z$ coefficient (in $A_T^{(1)}$) which is of similar magnitude
as the asymptotic value of $A_T^{(1)}$ (and thus corresponds to the pion). The other
intersection point, away from the $\omega$-axis, is a longitudinal vector
(when the $1/z$ coefficient of these modes is normalized to one, their
value at~$z=\pm\infty$ goes to zero as~$k\rightarrow 0$). Following the
figures, there is a
clear sense in which a state which starts out as a pion eventually
turns into a longitudinal vector, and vice versa, as $\mu_I$ is increased.
As discussed in \S\ref{pionvects}, this transition is smooth for any
non-zero momentum.

\begin{figure}[tb]
\vspace{4ex}
\begin{center}
\hbox{\vbox{\hbox{\includegraphics[width=.32\textwidth]{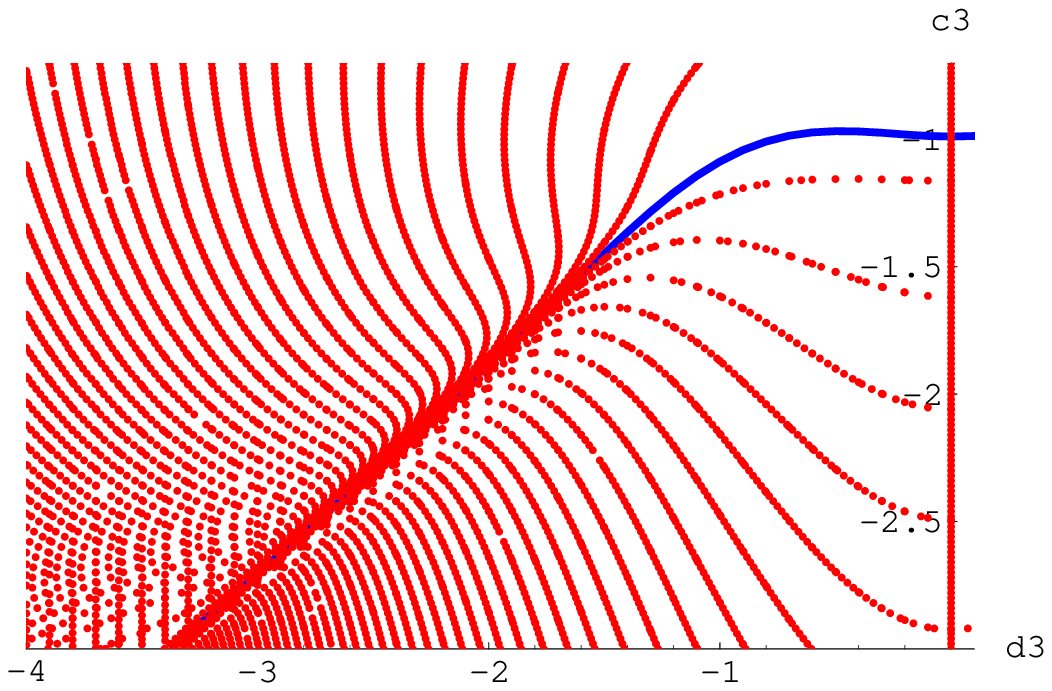}}\hbox{$\hspace{4em}\mu_I=1.57$}}
\vbox{\hbox{\includegraphics[width=.32\textwidth]{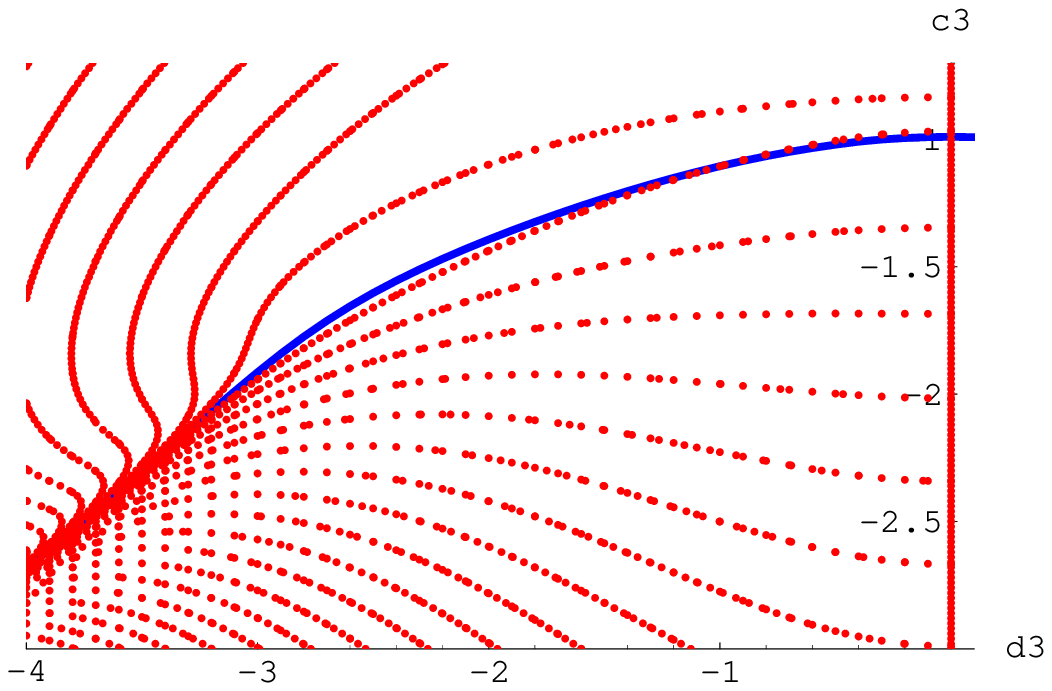}}\hbox{$\hspace{4em}\mu_I=2.10$}}
\vbox{\hbox{\includegraphics[width=.32\textwidth]{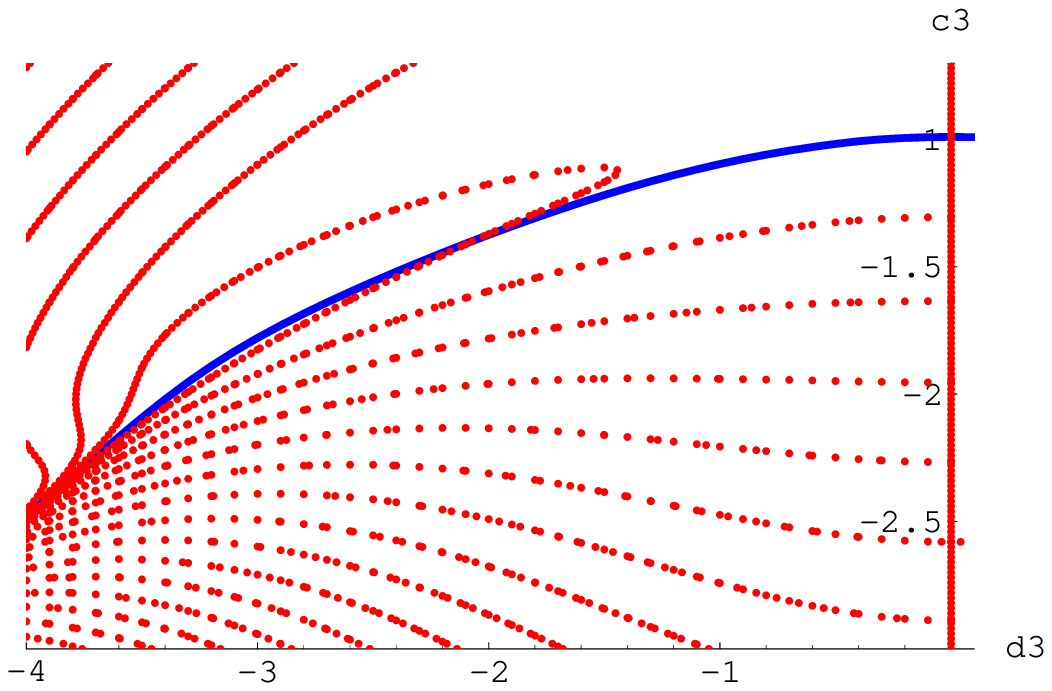}}\hbox{$\hspace{4em}\mu_I=2.20$}}}
\end{center}
\vspace{-4ex}
\caption{The location of solutions to the
  non-linear equations of motion~\protect\eqref{e:nonlinear}
  for the rho meson condensate, including the new ground state, in the
  same units as in figure~\protect\ref{condensate_fig}.
  Blue curves indicate the parameter
  values for which~$A_0^{(3)}(z=-\infty)$ satisfies the boundary
  condition, while red dots do the same for~$A_3^{(1)}(z=-\infty)$.\label{f:nonlinear}}
\end{figure}

For the construction of the rho meson condensate, a scan is required
through the space of $c_3$ and $d_3$ coefficients in the expansion
\begin{equation}
A_0^{(3)} = \mu_I\,\Big(\frac{1}{2} + \frac{c_3}{\pi z}\Big) + \ldots\,,\qquad
A_3^{(1)} = \frac{d_3}{z} + \ldots\,.
\end{equation}
Again we plot the curves at which~$A_0^{(3)}(z=-\infty)$
and~$A_3^{(1)}(z=-\infty)$ satisfy the boundary conditions, and look
for intersection points. The pion condensate solution~\mbox{$c_3=-1,\,d_3=0$}
satisfies~\eqref{e:nonlinear} for all values of~$\mu_I$; this is the
point on the vertical axis in
figure~\ref{f:nonlinear}. For~\mbox{$\mu_I>\mu_{\text{crit}}$} a new
intersection point develops.

\setlength{\bibsep}{1.5pt}


\begingroup\raggedright\endgroup

\end{document}